\documentclass{article}

\usepackage{arxiv}

\usepackage[utf8]{inputenc} 
\usepackage[T1]{fontenc}    
\usepackage{hyperref}       
\usepackage{booktabs}       
\usepackage{amsfonts}       
\usepackage{nicefrac}       
\usepackage{microtype}      
\usepackage{lipsum}
\usepackage{graphicx}
\graphicspath{ {./images/} }
\usepackage{natbib}
\usepackage{url}
\usepackage{graphicx,subcaption,bm}
\usepackage[utf8]{inputenc}
\usepackage{subcaption,cuted}	
\usepackage{amsthm}
\usepackage{amsmath}
\usepackage{mathtools}
\usepackage{amssymb}
\usepackage{changepage}
\usepackage{hyperref,cancel}
\usepackage{lipsum,listings,nccmath}
\usepackage{float,array,  longtable, ragged2e}
\usepackage{nomencl}
\usepackage{diagbox}
\usepackage[export]{adjustbox}
\usepackage{chngcntr}
 \usepackage{multirow}
 \usepackage{algorithm}
 \usepackage{algpseudocode}
 \usepackage{pdflscape,comment}
 \usepackage{multicol}
 \usepackage{svg}
\usepackage[nameinlink,noabbrev]{cleveref}
\usepackage{tikz}
\usepackage{algorithm}
\usepackage{algpseudocode}
\usepackage{enumitem}
\usepackage{tikz}
\usepackage{dsfont}
\usepackage[title]{appendix}
\usepackage{dsfont}
\usepackage{booktabs}

\title{Quantifying Exposure Information Uncertainty in Regional Risk Assessment}

\author{
Chenhao Wu \\
   Department of Civil and Environmental Engineering\\
  University of California\\
  Los Angeles, CA 90095 \\
  \texttt{chenhaowu@ucla.edu} \\
   \And
 Henry V. Burton\\
    Department of Civil and Environmental Engineering\\
  University of California\\
  Los Angeles, CA 90095 \\
  \texttt{hvburton@ucla.edu} \\
}

\begin{document}
\maketitle
\begin{abstract}
Exposure characterization in regional risk assessment aims to assign physical properties to the assets of interest so they can be associated with damage and loss functions. While this process has benefited from the growing availability of public infrastructure inventories, these datasets often lack the detailed attributes required for high-resolution risk assessment. Missing attributes are commonly inferred using predictive models or engineering-based rulesets. However, these imputations are inherently imperfect and can introduce bias and additional uncertainty in regional risk estimates. This study proposes a methodology to quantify the bias and uncertainty in regional risk assessment that arises from probabilistic exposure characterization. By integrating analytical and simulation-based approaches, the methodology decomposes the total uncertainty into contributions from incomplete exposure information as well as other sources, including hazard and damage characterization. This decomposition clarifies how bias and uncertainty associated with missing exposure information are generated and propagated through the risk assessment pipeline. The methodology is applied to both bridge-specific and regional risk assessments. A high-resolution bridge exposure inventory is developed using a data augmentation framework that combines publicly available information with machine learning and engineering-based imputation methods. 
\end{abstract}


\section{Introduction}
Exposure characterization plays a foundational role in regional risk assessment by identifying and describing physical (and sometimes human) assets that can be adversely affected by the hazard(s) of interest. It specifies what is at risk by compiling comprehensive inventory-scale information on these assets, which often include buildings, bridges, and other lifelines and critical facilities \citep{davis2024nist1310v1}. Each asset within a study region is described by a set of attributes that can capture its structural, non-structural, and other performance- and value-related characteristics. Examples of attributes include the year of construction, geometric properties, force-resisting system, and maintenance condition \citep{hazus}. These attributes collectively represent the essential information required to capture the asset behavior and potential consequences that may arise when subjected to some hazard-induced loading. Exposure characterization, therefore, establishes the link between the physical characteristics of an asset and the underlying mechanisms that drive hazard-induced consequences. From a mathematical modeling viewpoint, it is one of the key inputs used to relate hazard-induced loading intensity to expected damage states through fragility functions. Thus providing the basis for estimating direct economic loss \citep{chen2026high}, recovery durations \citep{wu2025modeling}, and other indirect impacts \citep{zhu2025macroeconomic}. 

In a regional risk assessment, the choice of modeling fidelity and resolution influences the attributes required to characterize individual assets and the effort needed to collect the corresponding exposure data. In this context, fidelity refers to the level of detail used to define an individual asset, whereas resolution refers to the extent to which multiple assets are represented uniquely or grouped based on their similarity \citep{dahal2025high}. Together, these choices influence the precision and type of information produced by the regional assessment. Regional risk can be generally quantified using component- and/or system-level assessments. A system-level assessment treats individual assets as holistic entities, without explicitly modeling their constituent components. In contrast, a component-level assessment requires more detailed information such as the quantities and configurations of the structural and non-structural elements in a building, or the number of columns, bearings, and other parts of a bridge. Component-level performance metrics can be aggregated to evaluate the system performance. Conversely, system-level performance metrics are inherently coarse and lack the granularity needed to infer component-level impacts, and thus represent a lower fidelity approach. 

A high-fidelity, high-resolution risk assessment requires substantial effort to collect the necessary exposure data that describes individual assets and their associated components. The process usually involves multiple data sources, each contributing complementary information related to physical properties. Prior research (e.g., \cite{anagnos2010development, lin2017stochastic, lochheaddata, kijewski2023validation}) has used a variety of data sources, including publicly available national inventories (e.g., the National Structure Inventory (NSI), the National Bridge Inventory (NBI), and the Homeland Infrastructure Foundation-Level Data (HIFLD)) and proprietary datasets. However, the exposure data from these sources are often generic and only defined at the system-level, thereby lacking the level of detail required for component-level modeling. To address these limitations, engineers perform field or virtual inspections \citep{anagnos2010development,pavic2020development} and review as-built drawings or photographs to supplement available exposure data. Moreover, attributes like construction details and maintenance and retrofit history may only be available through direct communication with local agencies. When such communication is not possible, design details must be inferred using proxy variables such as the year of construction, applicable design code, and code history. 

Given the substantial manual effort required to retrieve attributes not publicly available, predictive models, particularly those based on machine learning (ML), have become increasingly valuable. In this context, the task of compiling a complete inventory is framed as a missing data problem. Several prior studies have leveraged computer vision techniques to automate attribute extraction from satellite imagery, street-level imagery, and parcel footprints \citep{wang2021machine, ghione2022building, gomez2025automating, hamburger2025computer}.  However, the use of vision-based feature extraction assumes the availability of high-quality ground and aerial imagery, which is often impractical. ML models based on tabular data have also been used to impute missing attributes in building inventories \citep{song2026reliable,ghasemi2026integrated}. 

Existing imputation-based approaches have mainly focused on improving the accuracy performance of the predictive models. However, a few studies have investigated how exposure information availability (and lack thereof) propagates into downstream risk estimates \citep{gomez2022epistemic,yi2025impact,ghasemi2026integrated, song2026reliable}.  In this study, the term ``exposure information" is used to reflect the data that is used as input into the damage and loss steps of the risk assessment. We elaborate on this further in \autoref{sec2:concept}. Since this imputation is inherently imperfect, it can result in significant misrepresentation of asset attributes, and hence bias and add uncertainty to risk estimates. Prior studies have examined the effects of limited exposure information availability on building-specific structural response \citep{yi2025impact} and portfolio-scale (also building) loss estimation \citep{gomez2022epistemic,ghasemi2026integrated}. 

\begin{figure*}[h]
        \centering
     \includegraphics[width=.9\textwidth]{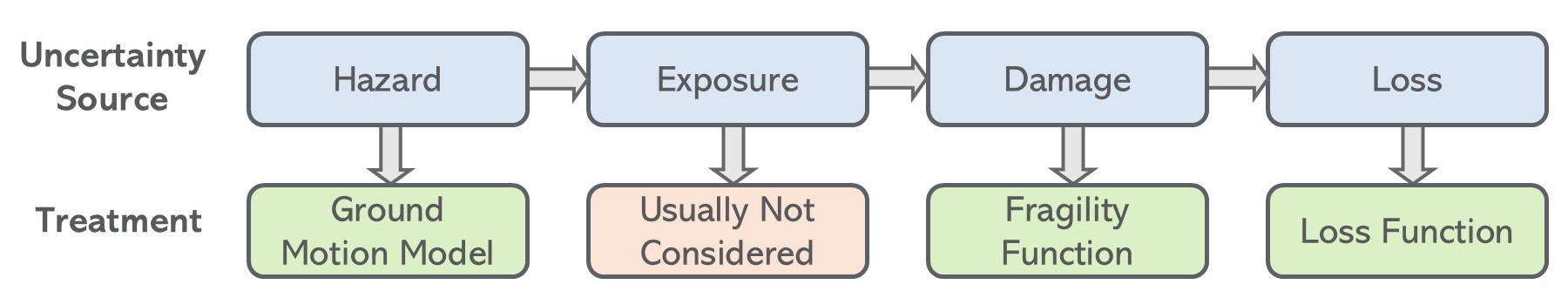}%
	\caption{Sources of uncertainty and their corresponding treatment in a regional risk assessment.}
	\label{fig:uncertainty_source}
\end{figure*}

\autoref{fig:uncertainty_source} summarizes the main sources of uncertainty in a standard regional risk assessment workflow, in which hazard, exposure, damage, and loss are evaluated in sequence. Considerable progress has been made in characterizing the uncertainties in three of these steps. At the hazard level, ground motion models coupled with spatial and spectral correlation algorithms quantify the joint distribution (including covariance) of shaking intensities across the considered sites \citep{boore2014nga,jayaram2009correlation}. In the damage assessment stage, fragility functions incorporate structure-specific sources of uncertainty, such as the variability in material properties and model parameter uncertainties, while also propagating the uncertainty from hazard characterization  \citep{hazus,chen2025second}. In the loss estimation stage, consequence functions are used to translate damage into uncertain impacts such as repair cost and recovery time \citep{chen2026high}. In contrast, uncertainty quantification in the exposure modeling step has received limited attention.
In some regional-scale risk assessments, assumptions have been made that there is perfect information about the underlying asset or the uncertainty is simply neglected \citep{mahsuli2013seismic, lin2017stochastic_part2, chen2026high, Wu_etal_regional_recovery}. These assumptions are particularly problematic in regional assessments, where incomplete inventory data often require probabilistic imputation to fill gaps, thereby introducing additional uncertainty into risk estimates. 

The preceding review highlights an important gap in the regional risk assessment workflow. Compared to other steps, there remains no systematic approach to quantifying the uncertainty in exposure characterization introduced by information imputation. Additionally, exposure inventory development has primarily centered on building portfolios. In contrast, critical infrastructure systems, such as bridge networks, have received comparatively less attention in both the literature and available toolkits.  To address this gap, this study develops a methodology to explicitly quantify and propagate the bias and uncertainty from exposure modeling within a regional risk assessment framework. The remainder of the article is organized as follows. The next section elaborates on the concept of ``exposure information uncertainty'', a type of epistemic uncertainty that emerges when imperfect imputation is used to characterize the considered assets. A unified methodology is proposed, combining analytical representations with Monte Carlo sampling, to quantify the associated uncertainty and track its propagation through the subsequent damage and loss assessment stages. The proposed methodology is demonstrated by using ML-based surrogate models to impute the missing information in a high-resolution seismic risk assessment of a bridge network. These models are informed by virtual inspections of more than 1,600 representative California bridges. The resulting inventory, combined with the proposed methodology, enables an evaluation of the impact of exposure information uncertainty at both the single-bridge and regional scales. 

\section{Exposure Information Uncertainty} \label{sec2:concept}
The sources of uncertainty in engineering design and assessment are often categorized as aleatory or epistemic \citep{der2009aleatory}. The former describes irreducible, inherent randomness, and the latter is associated with the lack of knowledge, and can be reduced if more observations are made. In probabilistic risk assessment, the loading intensity produced by the hazard event(s) of interest is a well-recognized example of aleatory uncertainty. Examples of epistemic uncertainty include the variability in structural material and geometric properties, imperfect representations of structural modeling \citep{liel2009incorporating,yin2010seismic}, model parameter estimation uncertainty due to limited data \citep{beck2012robustness,wu2024effects}, and variability in repair cost and time \citep{bradley2010accuracy}.

This study uses the term ``exposure information uncertainty'' to describe a form of epistemic variability that arises when imperfect imputations are used to infer missing infrastructure attributes in the exposure modeling step. Examples of these missing attributes include the lateral force-resisting system of a building and the abutment type in a bridge, which are typically not available in public datasets (e.g., the National Building Inventory and National Bridge Inventory) but are required for a risk assessment. To impute these attributes, engineering-based rulesets or data-driven surrogate models can be used to generate predictions with associated confidence levels. Regardless of the imputation strategy used, all such methods are inherently imperfect and can introduce additional epistemic uncertainty as well as potential bias due to misrepresentation.  This added uncertainty and bias propagate through the risk assessment pipeline and ultimately influence the downstream impact estimates (e.g., economic loss and recovery).

In many risk assessments, the asset attributes used as the basis of structural response simulation models or fragility functions are assumed deterministic. Under this assumption, there is no uncertainty associated with exposure information, and the predictive uncertainty of a risk quantity $Y$ (e.g., the damage state, repair cost, recovery duration) arises only from other sources i.e., hazard, damage, loss. In practice, however, the exposure information is always incomplete, and the imputation procedure produces a probability distribution over several possible classes. Let $\mathcal{E}$ denote a set of candidate exposure classes. The probability space is partitioned into $\mathcal{E} = \{\varepsilon _1,\dots,\varepsilon_i,\dots,\varepsilon_N\}$, with each $\varepsilon_i$ corresponding to a candidate exposure class defined by a particular combination of asset attributes. 
Within each exposure class, the exposure information is fixed, and the uncertainty from the remaining sources still persists. Samples of the outcome $Y$ can be generated according to the conditional distribution $p(Y|\varepsilon_i)$. 

Because multiple exposure classes now exist in the probability space, the total predictive uncertainty of $Y$ increases, compared to the case where there is a single class. This increase can be formally described using the law of total variance, which isolates the contribution of exposure information uncertainty to the total variance of $Y$:
\begin{equation} \label{eq:tot_var}
    \mathrm{Var}(Y)
    = \underbrace{\mathbb{E}\!\left[\mathrm{Var}(Y\mid \mathcal{E})\right]}_{\mathrm{Var}^{\text{base}}(Y)}
    + \underbrace{\mathrm{Var}\!\left(\mathbb{E}[Y\mid \mathcal{E}]\right)}_{\mathrm{Var}^{\text{e}}(Y)} .
\end{equation}
The first term, $\mathrm{Var}^{\text{base}}(Y)$, is referred to as the baseline uncertainty. This term represents the uncertainty that would remain even if the correct class is known. In other words, once a class $\varepsilon_i$ is fixed, the uncertainty associated with exposure information disappears, and the remaining uncertainty in $Y$ is solely from the other sources. Note that the remaining uncertainty includes all other sources not explicitly conditioned upon, such as in the ground shaking intensity of (aleatory) or model parameter definition (epistemic). 

The second term, $\mathrm{Var}^{\text{e}}(Y)$, represents the additional variability arising from uncertainty in the exposure class. This term measures the spread of the conditional means $\mathbb{E}[Y\mid \varepsilon_i]$ across the candidate classes. In the limiting case where the true class is known, the variability across conditional means disappears and $\mathrm{Var}^{\text{e}}(Y)$ becomes zero, and the total predictive uncertainty reduces to the baseline variance. 

\begin{figure*}[h] 
        \centering
     \includegraphics[width=.65\textwidth]{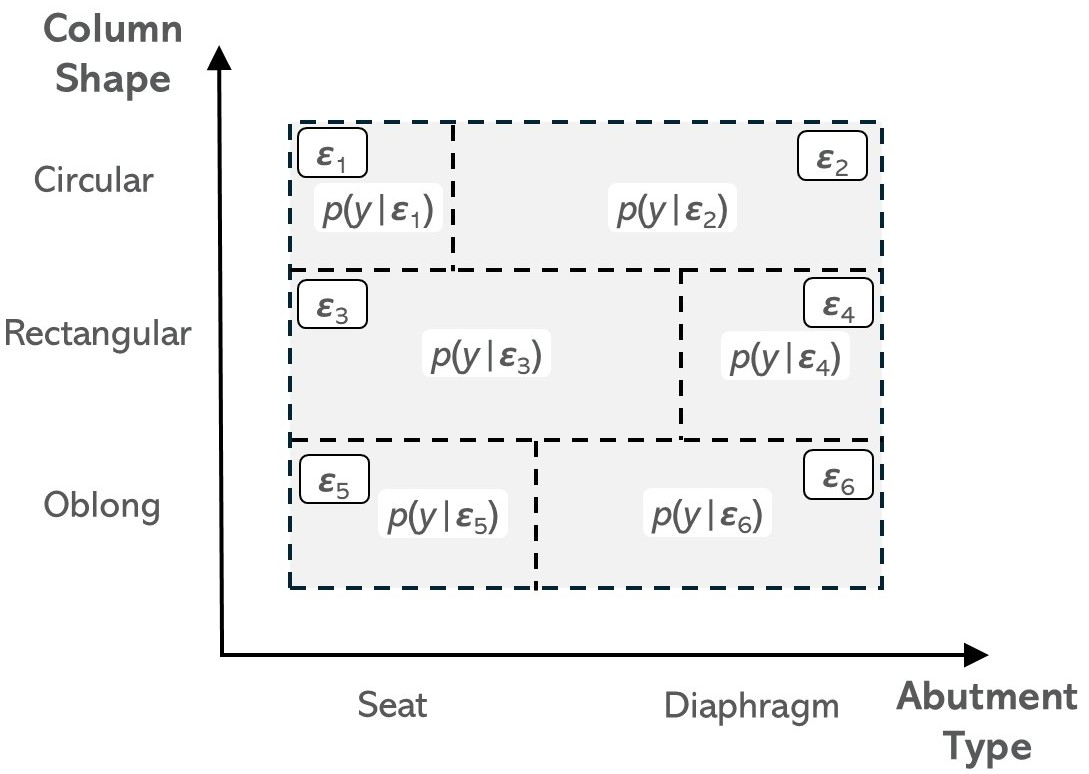}%
	\caption{Schematic illustration of probability space partition due to the existence of candidate exposure classes.}
	\label{fig:prob_space_partition}
\end{figure*}

This concept can be graphically illustrated in \autoref{fig:prob_space_partition}. Consider a surrogate model used to predict two missing bridge attributes: the abutment type and column shape. Because the model is uncertain about the true value of these attributes, it assigns probabilities to each possible category. The resulting probability space is then partitioned into six candidate exposure classes, each of which represents a unique combination of the possible attribute values. Within each class, samples of the outcome $Y$ can be generated according to the corresponding within-class probability, $p(y|\boldsymbol{\varepsilon}_i)$, where $\boldsymbol{\varepsilon}_i$ represent a joint observation of abutment type and column shape. The additional uncertainty resulting from the imperfect imputation is captured by the variability of within-class means (i.e., the $\mathrm{Var}^\text{e}(Y)$ term in \autoref{eq:tot_var}). Note that this variance decomposition has been used in prior risk and reliability studies (e.g., \cite{jeon2026ensemble}) to isolate a particular source of uncertainty from the total variance. 

\section{Quantification and Propagation of Exposure Information Uncertainty} \label{sec3:quantify}
Consider a bridge that is characterized by two distinct sets of attributes. The first set,  $\bm{X} = \{X_1,\dots, X_p\}$, contains readily available attributes, such as design era, number of spans, and span length. These can typically be retrieved from a public inventory with well-documented tabular data. The second set, $\bm{\tilde{X}} = \{\tilde{X}_1,\dots,\tilde{X}_q\}$, encompasses attributes that are not directly obtainable from public databases, such as the abutment type, bent type, and number of columns. Together, the concatenated set $\{\bm{X},\bm{\tilde{X}}\}$ constitutes a complete inventory that is required for regional risk assessment. 

As discussed in the preceding section, imputation methods can be used to estimate $\bm{\tilde{X}}$ based on the available $X$. Given the inherent imperfection in this imputation, the estimated $\bm{\tilde{X}}$ contains misrepresentations relative to the ground-truth. These errors will propagate through the subsequent damage and loss estimation stages of the risk assessment workflow (\autoref{fig:uncertainty_source}), and ultimately introduce bias and additional uncertainty into the downstream risk metrics. This section presents a methodology for systemically quantifying and tracking this type of uncertainty through the risk assessment pipeline. Note that the formulations developed in this section assumes $\bm{\tilde{X}}$ consists of categorical variables. However, they can also be applied to continuous attributes through the appropriate discretization.

\subsection{Probabilistic prediction of attribute values using a multi-class classifier}
Suppose the prediction task is to estimate the probability that a missing attribute belongs to class $k\in\{1,\dots,K\}$ given a known input feature $\bm{X}$, denoted as $P(\tilde{X}=k \mid \bm{X})$. For many commonly used classification algorithms, such as multinomial logistic regression, neural networks, and boosting-based methods, the model first maps \(\bm{X}\) to a set of class-specific latent score functions $\{f_k(\bm{X})\}_{k=1}^K$, where the functional form of $f_k(\bm{X})$ depends on the selected classifier. These scores are then transformed into a normalized probability vector through the softmax function:
\begin{equation} \label{eq:softmax}
    P(\tilde{X}=k|\bm{X}) = \frac{\exp(f_k(\bm{X}))}{\sum_{j=1}^K \exp({f_j(\bm{X})})}, \,\,k=1,\dots,K.
\end{equation}
Under this formulation, different classification algorithms primarily differ in how $f_k(\bm{X})$ is estimated, while the final class probability distribution is estimated by mapping the resulting scores into a softmax function. This perspective provides a unified representation for a broad class of score-based classifiers. Other classification algorithms, such as decision trees and random forests, do not explicitly estimate class-specific latent scores in this form, and their class probabilities are derived through alternative mechanisms such as empirical frequencies. These approaches are not extensively discussed here. 

A well-known limitation of softmax probabilities is that they are often poorly calibrated, meaning the predicted probabilities do not reflect empirical frequencies and hence cannot be interpreted as probabilities of occurrence \citep{niculescu2005predicting}. In particular, many modern models tend to produce overconfident predictions. To address this limitation, post-hoc calibrations are used to align predicted probabilities with empirical frequencies. Common calibration techniques include temperature scaling, Platt scaling, and isotonic regression \citep{guo2017calibration}. In the temperature scaling approach, the scores are rescaled before applying the softmax function. Specifically, given the score functions $\{f_k(\bm{X})\}_{k=1}^K$, the calibrated probabilities are defined as: 
\begin{equation} \label{eq:softmax_calibrated}
    P(\tilde{X}=k|\bm{X}) = \frac{\exp(\frac{f_k(\bm{X})}{T})}{\sum_{j=1}^K \exp({\frac{f_k(\bm{X})}{T}})}, \,\,k=1,\dots,K,
\end{equation}
where $T$ is a temperature parameter that is learned by minimizing the negative log-likelihood on a validation set (a portion of data held out from the training set). When $T>1$, the predicted probabilities are softened (reducing overconfidence), whereas $T<1$ sharpens the probabilities. 

\subsection{Propagating exposure uncertainty to damage estimation stage} \label{sec:uncertainty_damage}
In regional risk assessment, it is generally infeasible to perform explicit nonlinear structural analyses for every asset in a network to evaluate damage. Instead, damage is typically estimated by sampling from available fragility functions, which serve as a surrogate for explicit structural analysis \citep{heresi2023regional,chen2026high}.

Consider a case where a bridge is assigned to class $i$ based on attributes derived from both observed data and imputed values. Conditional on that class assignment, a set of pre-established fragility functions can be associated with the bridge. The fragility function for the $i^{\text{th}}$ class is expressed as:
\begin{equation} \label{eq:group_based_fragility}
    P_i(DS \ge ds_k \mid IM=im) = \Phi\left( \frac{\ln(im) - \ln(\theta_{k,i})}{\beta_{k,i}}\right),
\end{equation}
where $\Phi(\cdot)$ is the standard normal cumulative distribution function, $\theta_{k,i}$ is the median intensity measure corresponding to damage state threshold $ds_k$ for class $i$, and $\beta_{k,i}$ is the logarithmic standard deviation.

The corresponding in-state damage probability for class $i$ conditioned on the ground motion intensity measure $(IM=im)$ is then given by:
\begin{equation} \label{eq:in_state_group_fragility}
        p_{k,i}(im) = 
\begin{cases}
1 - P_i(DS \ge ds_1 \mid IM=im) & \text{if } k = 0, \\[2pt]
P_i(DS \ge ds_k \mid IM=im) - P_i(DS \ge ds_{k+1} \mid IM=im) & \text{if } 1\le k < N_{DS}, \\[2pt]
P_i(DS \ge ds_{N_{DS}} \mid IM=im) & \text{if } k = N_{DS}.
\end{cases}
\end{equation}

As discussed earlier, because of misrepresentation in the surrogate-based attribute imputation model, the inferred bridge class is no longer deterministic. Instead, it is represented by a probability distribution $\bm{\pi}=\{ \pi_1,\dots,\pi_N\}$, with $\pi_i=P(\varepsilon_i|\bm{X})$.  The uncertainty in the attribute prediction propagates to damage assessment and may introduce both bias and additional uncertainty in the predicted damage probabilities. Specifically, the mean in-state damage probability is obtained  by weighting the class-specific probabilities by $\bm{\pi}$:
\begin{equation} \label{eq:mean_damage_prob}
    \bar{p}_{k}(im) = \sum_{i=1}^{N} p_{k,i}(im)\pi_i.
\end{equation}

This mixture-based damage probability may deviate from the damage probability associated with the ground-truth class. Denoting the ground-truth bridge class as $i^{\text{True}}$, the bridge-level bias in the in-state damage probability can then be defined as:
\begin{equation} \label{eq:bias_in_state_prob}
    \text{Bias}\!\left(p_k(im)\right) = \bar{p}_{k}(im) - p_{k,i^{\text{True}}}(im).
\end{equation}

This bias may lead to an overestimation of the damage severity when a bridge is misclassified into a more vulnerable class (e.g., a diaphragm-type abutment misclassified as a seat-type). Conversely, the level of damage may be underestimated when the bridge is incorrectly assigned into a less vulnerable class (e.g., a single-column bent misclassified as a multi-column bent). In either case, exposure information uncertainty arises because the resulting in-state damage probability is itself uncertain under class ambiguity. This term can be quantified by the standard deviation of the in-state damage probability across plausible classes, given by:
\begin{equation} \label{eq:epis_in_state_prob}
\text{sd}^\text{M}\!\left(p_k(im)\right) =
    \sqrt{\sum_{i=1}^{N} \left[p_{k,i}(im) - \bar{p}_{k}(im)\right]^2\,\pi_i }.
\end{equation}
Note that this derivation aligns with $\sqrt{\mathrm{Var^\text{e}(Y)}}$ in \autoref{eq:tot_var} by treating $Y=\mathds{1}(DS=ds_k)$.

\subsection{Propagating exposure uncertainty to loss estimation stage} \label{sec:uncertainty_loss}
Conditioned on the ground motion intensity measure, the repair cost for a single bridge can be expressed as \citep{mackie2010post,chen2026high}: 

\begin{equation} \label{eq:tot_loss}
    L(im) =  RP(im) + RE(im),
\end{equation}
where the replacement component, $RP(im)$, is calculated as the replacement cost ($RPC$) weighted by the probability of collapse: 

\begin{equation} \label{eq:loss_collapse}
    RP(im)=RPC \cdot p_c(im).
\end{equation}

The reparable component, $RE(im)$, is calculated by aggregating contributions from individual components over their associated damage states, expressed as:

\begin{equation} \label{eq:loss_repairable}
    RE(im) = \sum_j\left[\sum_k C_{j,k}\cdot \left(\frac{p_{j,k}(im)}{\sum_kp_{j,k(im)} + p_{j,0}(im)}\right) \right] \cdot(1-p_c(im)),
\end{equation}

where $p_{j,0}(im)$ denotes the probability of no damage for the $j^{th}$ component conditioned on the intensity measure $im$, and $C_{j,k}$ denotes the total repair cost for that component under the $k^{th}$ damage state. The $C_{j,k}$ term can be further decomposed as the product of the quantity of that component and its unit repair cost, i.e., $C_{j,k} = n_j\cdot UC_{j,k}$.

Bias and exposure information uncertainty arise in the loss estimation stage through two primary mechanisms. First, the bias and uncertainty from the damage estimation (i.e., \autoref{eq:bias_in_state_prob} and \autoref{eq:epis_in_state_prob}) propagate directly into the loss calculation. In particular, the bias and uncertainty in the estimated in-state probabilities ($p_{j,k}(im)$ and $p_c(im)$) will affect the loss predictions. Second, additional bias and epistemic uncertainty are introduced through the imputation of component quantities ($n_j$), which directly influence repair cost estimates.

Analytical expressions can, in principle, be derived to quantify the propagation of these uncertainties. However, such derivations can become cumbersome due to the interaction among multiple uncertainty sources. Moreover, these closed-form formulations are often difficult to generalize when additional sources are introduced. To address this, a unified Monte Carlo sampling scheme is adopted to quantify the combined effects of exposure information and other sources of uncertainty. The details of this approach is described in Algorithm~\ref{alg:mc_model_info}, which accounts for the uncertainty in the imputed exposure class, along with the variability that originates in the damage and loss assessment stages. In each simulation, an exposure class is sampled, corresponding to one realization of the unknown attributes. The resulting loss is evaluated based on that realization to produce a system-level outcome. Repeating this process yields a set of simulated outcomes that capture both sources of uncertainty. 

The simulated outcomes are used to estimate the total uncertainty and its decomposition into two components: the baseline variance, computed as the empirical mean of the class-conditional variances, and the exposure information uncertainty, computed as the empirical variance of class-conditional means. This procedure enables uncertainty from damage, loss, and exposure information to be sampled in a consistent manner, while allowing the latter to be isolated from the total uncertainty in a subsequent step. 
 
\begin{algorithm}[htbp]
\caption{Sampling scheme for estimating the baseline and exposure information uncertainty}
\label{alg:mc_model_info}
\begin{algorithmic}[1]

\State Let $\{\varepsilon_1,\dots,\varepsilon_i,\dots,\varepsilon_N\}$ denote the set of possible exposure classes.
\State Specify the total number of simulations $R$.

\For{$r = 1,\dots,R$}
    \State Sample a class $\varepsilon^{(r)}$ from its probabilistic distribution.
    \State Form a complete inventory of component quantities based on both the imputed and known attributes.
    \State Sample component damage from the assigned fragility functions (\autoref{eq:in_state_group_fragility}).
    \State Sample repair cost based on the assigned loss functions and Equations~\eqref{eq:tot_loss}--\eqref{eq:loss_repairable}, denoted by $y^{(r)}$.
\EndFor

\For{$i = 1,\dots,N$}
    \State Compute the number of simulated losses assigned to class $\varepsilon_i$:
    $$
    N_i = \sum_{r=1}^{R} \mathbf{1}\!\left(\varepsilon^{(r)} = \varepsilon_i\right)
    $$
    
    \State Compute the empirical probability of class $\varepsilon_i$:
    $$
    \hat{\pi}_i = \frac{N_i}{R}
    $$
    
    \State Compute the class-conditional sample mean:
    $$
    \hat{\mu}_i =
    \frac{1}{N_i}
    \sum_{r:\,\varepsilon^{(r)}=\varepsilon_i} y^{(r)}
    $$
    
    \State Compute the class-conditional sample variance:
    $$
    \hat{\sigma}_i^2 =
    \frac{1}{N_i-1}
    \sum_{r:\,\varepsilon^{(r)}=\varepsilon_i}
    \left(y^{(r)}-\hat{\mu}_i\right)^2
    $$
\EndFor

\State Compute the overall sample mean:
$$
\hat{\mu} = \sum_{i=1}^{N} \hat{\pi}_i\hat{\mu}_i
$$

\State Estimate the baseline variance:
\begin{equation}  \label{eq:sample:baseline_var}
    \hat{\sigma}_{\mathrm{Base}}^2
=
\sum_{i=1}^{N} \hat{\pi}_i\hat{\sigma}_i^2
\end{equation}

\State Estimate the exposure information variance:
\begin{equation} \label{eq:sample:model_info_var}
\hat{\sigma}_{\mathrm{e}}^2
=
\sum_{i=1}^{N}
\hat{\pi}_i
\left(\hat{\mu}_i-\hat{\mu}\right)^2    
\end{equation}

\State The sample total variance is the sum of the baseline and exposure information variances:
$$
\hat{\sigma}_{\mathrm{Total}}^2
=
\hat{\sigma}_{\mathrm{Base}}^2 + \hat{\sigma}_{\mathrm{e}}^2
$$

\end{algorithmic}
\end{algorithm}

\subsection{Propagating exposure uncertainty to regional loss}
The regional economic loss ($RL$) is defined as the sum of individual bridge losses: 
\begin{equation} \label{eq:RL_var_decompose}
    RL = \sum_{b=1}^B L_b, 
\end{equation}
where $L_b$ represents loss for the $b^{\text{th}}$ bridge within the network. Applying the same variance decomposition introduced in \autoref{eq:tot_var} to $RL$:
\begin{equation} \label{eq:total_var_regional}
    \mathrm{Var}(RL)
    = \underbrace{\mathbb{E}\!\left[\mathrm{Var}(RL\mid \mathcal{E})\right]}_{\mathrm{Var}^{\text{base}}(RL)}
    + \underbrace{\mathrm{Var}\!\left(\mathbb{E}[RL\mid \mathcal{E}]\right)}_{\mathrm{Var}^{\text{e}}(RL)}.
\end{equation}

The decomposition of regional losses (\autoref{eq:total_var_regional}) builds on the individual bridge-level decomposition (Algorithm~\ref{alg:mc_model_info}). However, several important considerations arise when extending the variance decomposition to the regional scale: (1) The $\mathcal{E}$ here no longer refers to the exposure class associated with a single bridge. Instead, it represents the joint distribution of exposure classes across bridges in the network. (2) Because the regional loss is the sum of individual bridge losses, the dependence structure among bridge losses must be explicitly considered. Therefore, the variance of $RL$ depends not only on the variance for each $L_b$, but also on their covariance terms. 

Fortunately, this decomposition can still be evaluated using Monte Carlo simulations, without requiring analytical derivations of the full variance-covariance structure. By repeatedly sampling from the joint distribution of bridge classes and their associated losses, both the baseline and exposure information uncertainty can be estimated in a consistent manner. The detailed sampling procedure is provided in \autoref{app:sampling_scheme}.

\section{Imputation of Exposure Attributes} \label{sec:MLbridge}

This section describes the imputation of missing exposure attributes that are required for regional risk assessments of bridge networks. The NBI \citep{NBI2025} is used as the primary source of bridge inventory data, where several attributes necessary for the risk analysis are not directly reported and must be inferred.

As discussed in the Introduction, the choice of model fidelity and resolution influences the required input attributes and precision of the performance metrics, while also affecting the computational cost of regional-scale simulations. In the context of bridge risk assessment, two representative levels of modeling approaches are considered: the HAZUS-based, system-level approach \citep{hazus}, and the higher fidelity component-level approach \citep{chen2025second}. The system-level approach characterizes damage and loss at the level of the entire bridge, whereas the component-level assessment explicitly represents the response and damage to individual elements. As a result, the HAZUS-based approach is generally associated with lower modeling fidelity compared to the component-level assessment, but also requires fewer detailed input attributes.

While higher-fidelity modeling approaches are desirable, they typically require more granular information to characterize individual assets. \autoref{tab:bridge_attributes_comparison} summarizes the bridge attributes required by the HAZUS- and component-level approaches, respectively. The latter relies on detailed, component-specific information, such as the quantities and configurations of individual structural elements. However, the data available in the NBI are limited under both fidelity settings. Although the NBI provides a standardized and comprehensive tabular database with bridge inventory, condition, and maintenance information across the United States, it does not include many of the attributes required for higher fidelity assessments.

To address this gap, this section focuses on the imputation of missing bridge attributes needed for component-level assessment. A hybrid methodology is introduced that integrates ML-based surrogate models with engineering-based rulesets to augment incomplete bridge data. To support this effort, virtual inspections are conducted for a set of representative California bridges, resulting in a curated inventory that serves as the basis for training and testing the ML-based surrogate models. Finally, the model development process is described, including training, hyperparameter optimization, and performance evaluation with an automated workflow.

\begin{table}[H]
\centering
\caption{Summary of bridge attributes required for the system- and component-based approaches and their availability in NBI.}
\label{tab:bridge_attributes_comparison} 
\renewcommand{\arraystretch}{1.15}
\setlength{\tabcolsep}{8pt}
\begin{tabular}{>{\raggedright\arraybackslash}p{6.3cm}|c|c|c}
\hline
\multicolumn{1}{c|}{\multirow{2}{*}{Attribute}} 
& \multicolumn{2}{c|}{Required by} 
& \multicolumn{1}{c}{\multirow{2}{*}{NBI Availability}} \\ \cline{2-3}
& System-level & Component-level & \\ \hline

Year built 
& \multirow{4}{*}{Yes} 
& \multirow{3}{*}{Yes} 
& \multirow{3}{*}{Yes}  \\

Number of spans 
&  
&  
&  \\

Superstructure material 
&  
&  
& \\ \cline{1-1} \cline{3-4}

Span continuity 
&  
& No 
& \\ \cline{1-1} \cline{3-3}

Bent type 
&  
& \multirow{5}{*}{Yes} 
& \multirow{5}{*}{No} \\ \cline{1-2}

Abutment type 
& \multirow{4}{*}{No} 
&  
& \\

Column shape 
&  
&  
& \\

Number of columns per bent 
&  
&  
& \\

Total number of columns, pier walls, joint seals, and bearings 
&  
&  
& \\ \hline

\end{tabular}
\end{table}

\subsection{Framework for augmenting exposure data}
\begin{figure*}[h] 
        \centering
     \includegraphics[width=.8\textwidth]{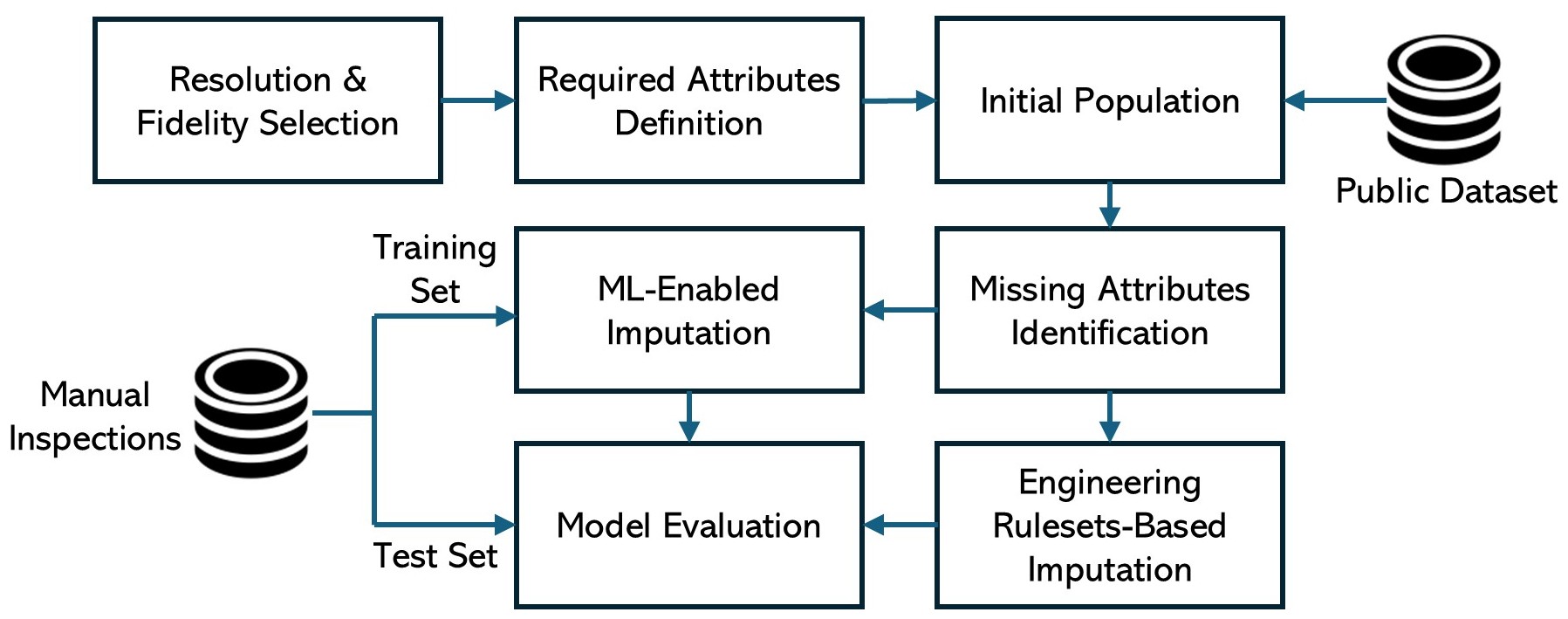}%
	\caption{Framework for augmenting bridge exposure inventory for regional risk assessment. Some of the terminologies are adopted from \cite{kijewski2023validation}.}
	\label{fig:framework_augmentation}
\end{figure*}

\autoref{fig:framework_augmentation} presents the framework for augmenting bridge exposure inventory data for regional risk assessment. The process begins by defining the fidelity and resolution used in the assessment. Based on the selected fidelity and resolution, the necessary bridge attributes are identified (\autoref{tab:bridge_attributes_comparison}). In the first stage of the exposure data acquisition, the NBI serves as the primary source of information to populate the relevant fields, including the bridge construction year and material and basic geometric properties. By comparing the initially populated dataset with the required attribute set, the missing fields can then be identified.

A hybrid imputation method is used to infer the missing fields, combining surrogate- and engineering ruleset-based imputation. This distinction is based on whether a specific attribute can be observed during an inspection. Observable attributes, such as bent type, abutment type, and number of columns, can be used to compile the data which are subsequently used to train and test the surrogate model. In contrast, hidden bridge attributes, such as foundation characteristics and other features that are difficult to identify from a virtual inspection (e.g., due to limited visibility, restricted viewpoint, and/or access) are inferred using engineering rulesets. In this study, the following attributes in \autoref{tab:bridge_attributes} are imputed using and ML-based surrogate model: bent type, abutment type, column shape, and number of columns. The details of this approach are described in subsequent subsections.  The remaining attributes are inferred through engineering-based rulesets, as detailed in the main text and supplementary material of \cite{Wu_etal_regional_recovery}. 

A complete bridge exposure inventory is then assembled through the integration of the observed and inferred inputs, with a probability distribution associated with the imputed attributes. This complete exposure inventory, together with the corresponding uncertainty information, is subsequently used as input for the probabilistic regional damage and loss assessments. 

Note that the framework shown in \autoref{fig:framework_augmentation} is not entirely new. Previous studies on inventory generation have used similar workflows, including data collection, missing field imputation, model training, and data fusion \citep{wang2021machine,kijewski2023validation,lochheaddata}. One of the key contributions of this work is the tailored design of these steps for bridge exposure data. 

\subsection{Data collection}
This subsection describes the virtual inspection and data collection efforts used to develop the ML models for bridges with a reinforced concrete (RC) bent. Two datasets were collected: the first comprises bridges from across California, and the second only includes bridges located in the City of Los Angeles (LA). The California statewide dataset serves as the training set, with the goal of capturing a broader range of bridge characteristics and improving model generalizability. The City of LA bridge inventory serves as the test set and is further used as the testbed for the subsequent regional risk assessment.
To ensure a balanced representation of bridge groups in the training set, we adopted a stratified sampling approach. The sampling procedure consists of three steps. First,  non-vehicular bridges (e.g., pedestrian bridges) are excluded from the baseline inventory. Then, the remaining bridges are stratified based on the following NBI attributes: design era, span class, and superstructure material. Lastly, the virtual inspection was performed for a random subset of the bridges in each stratum.

\begin{figure*}[h]   
 \centering
  \begin{subfigure}[h]{.45\textwidth}
	\centering
	\includegraphics[width=\textwidth]
    {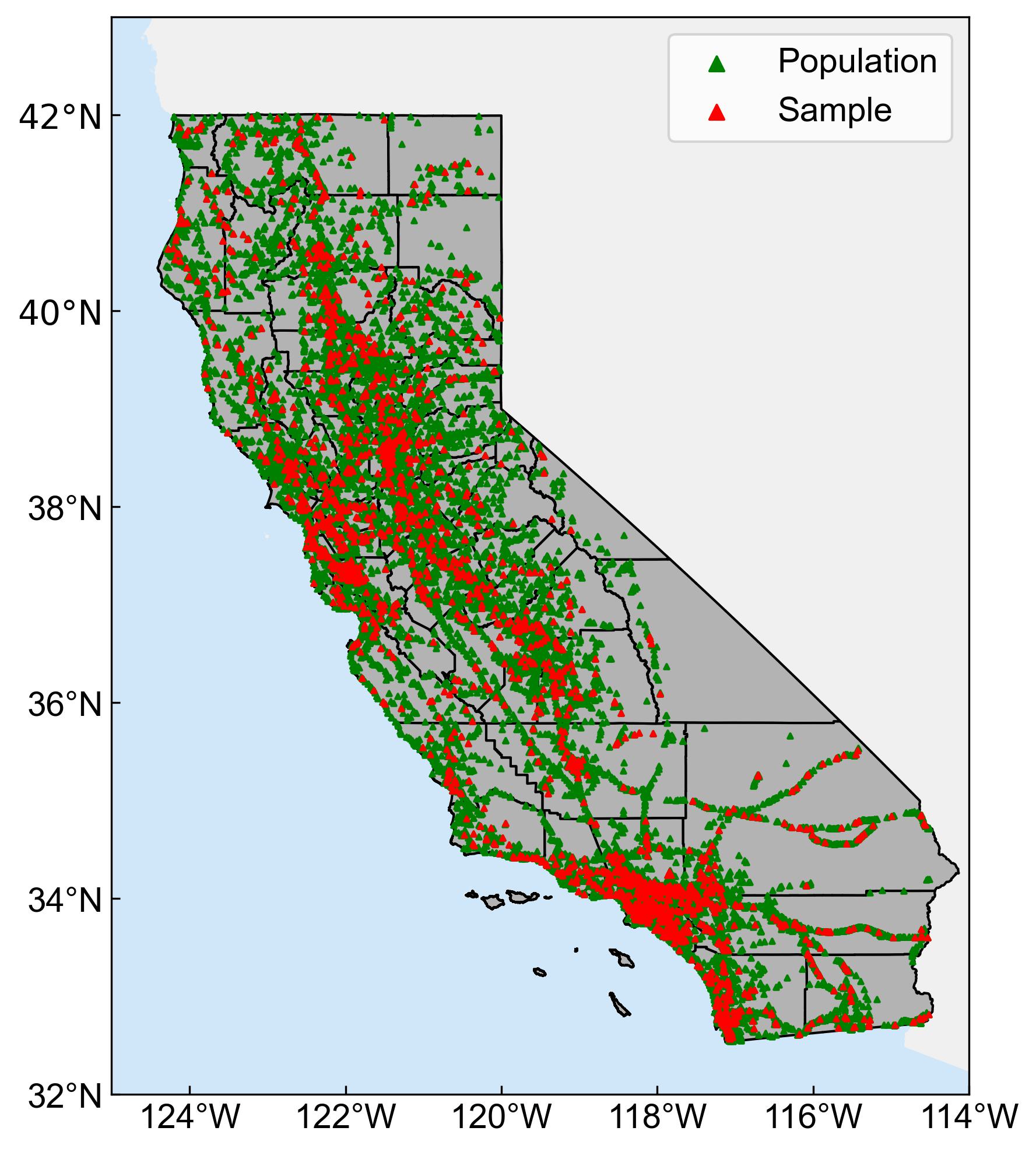}
	\caption{}
    \label{fig:dots_inspectedbridges}
\end{subfigure}
\begin{subfigure}[h]{.52\textwidth}
 	\centering
 	\includegraphics[width=\textwidth]{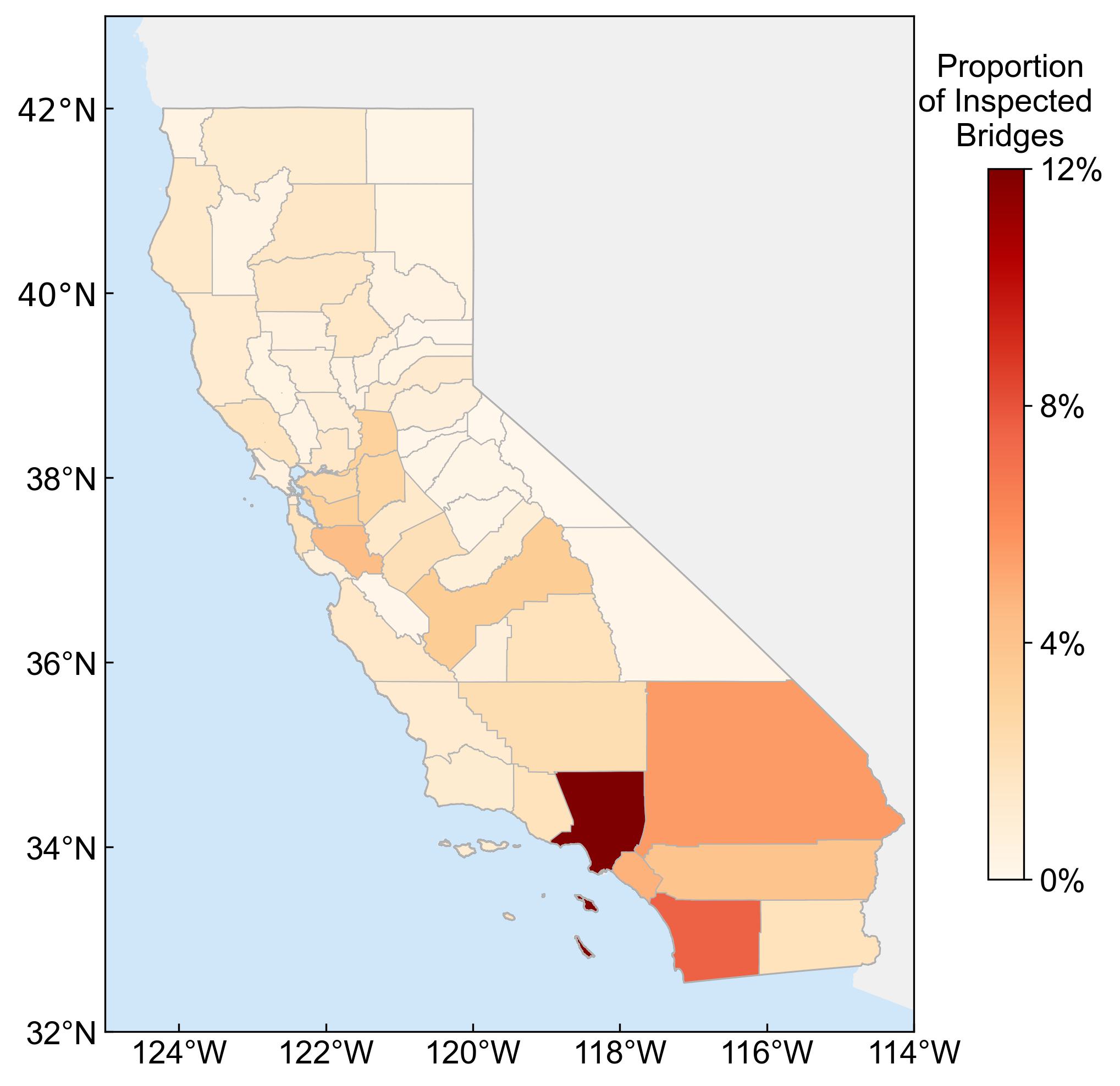}
 	\caption{}
    \label{fig:proportion_inspectedbridges}
 \end{subfigure}
\caption{California maps showing the: (a) spatial distribution of the complete and sampled bridge inventory and (b) the county-level proportion of bridges in the inspected sample.}
\label{figs:inspectedbridges}
\end{figure*}

\begin{figure*}[h] 
        \centering
     \includegraphics[width=.98\textwidth]{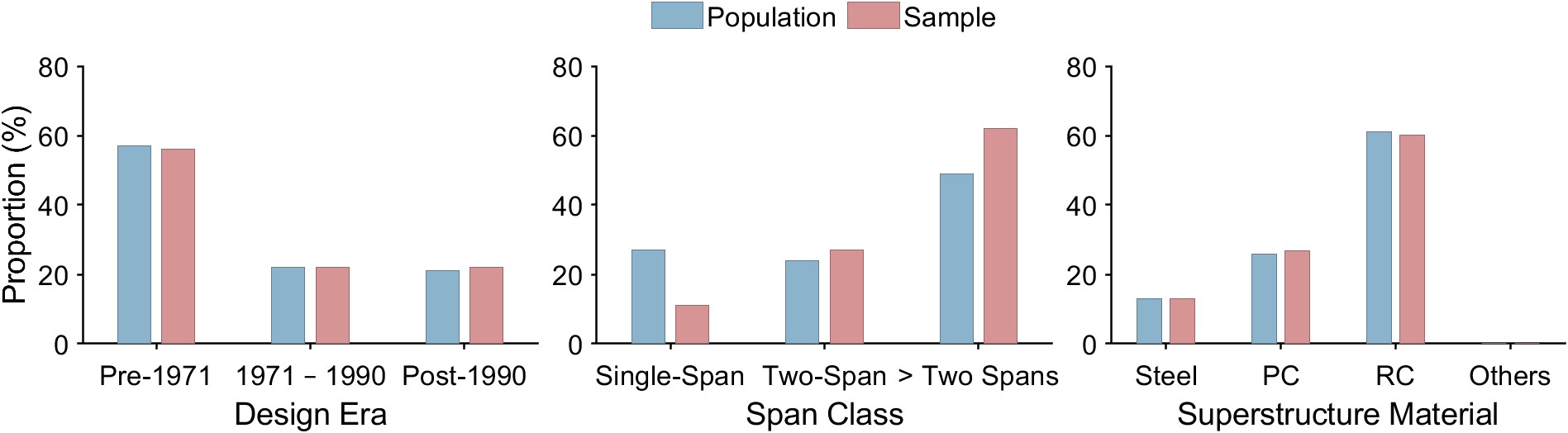}%
	\caption{Distribution of NBI attributes for the sampled and complete bridge inventory
    (RC: reinforced concrete. PC: precast concrete).}
	\label{fig:bar_comparison}
\end{figure*}

We selected 8\% of the California inventory for virtual inspection, corresponding to a total of 1,648 bridges. \autoref{figs:inspectedbridges} shows the spatial distribution of the sampled bridges relative to the full population, as well as the number of samples in each county. \autoref{fig:bar_comparison} compares the distributions of the three NBI attributes between the sampled bridges and the overall bridge population.

\begin{figure*}[h] 
        \centering
     \includegraphics[width=.95\textwidth]{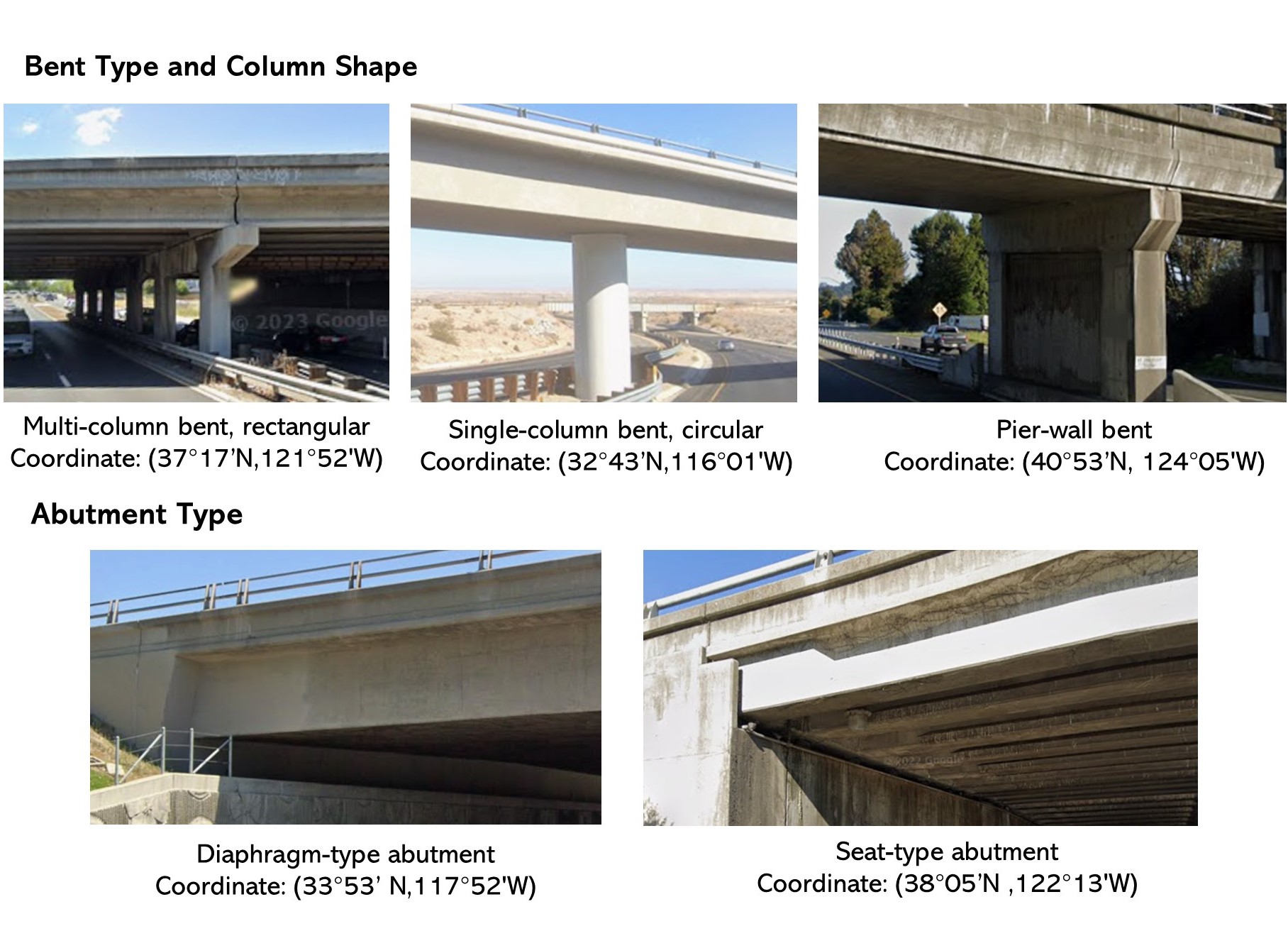}%
	\caption{Examples of bridge attributes identified during the virtual inspections (images extracted from Google Street View, \cite{GSV2025}).}
	\label{fig:inspection}
\end{figure*}

\autoref{fig:inspection} illustrates how bridge attributes are identified during the virtual inspections. The following four bent types are predefined based on their known effect on bridge performance: multi-column bent, single-column bent, pier-wall bent, and no-bent. The no-bent case applies only to single-span bridges and can be determined during the initial population stage based on the number of spans reported from NBI. For column bents, three column cross-sectional shapes are predetermined: rectangular, circular, and oblong. Two abutment types are also distinguished: the diaphragm-type and the seat-type. In a diaphragm-type abutment, the deck is monotonically cast into the abutment without a construction gap. In a seat-type abutment, girders are extended to the abutment by sitting on the abutment seat. In this case, distinct horizontal and vertical gaps are visible between the girder edge and the shear key. 

\subsection{Model training and performance evaluation} \label{sec:modeltrainingevaluation}
This subsection describes the use of ML models to probabilistically impute bridge attributes based on the data collected through virtual inspections. Four target attributes are considered in the prediction task: bent type, abutment type, column shape, and the number of columns per bent. The possible categories for these four attributes are listed in \autoref{tab:bridge_attributes}.

A total of nine features were selected from the NBI as input features for the ML model development. The geometric features include the deck area, deck width, span length, and presence of an irregularity. The design-related features include the design era, number of spans, superstructure material, and superstructure design. The traffic-related feature is represented by average daily traffic. The selection of these features was guided by engineering domain knowledge and their availability in the NBI.

\begin{figure*}[h] 
        \centering
     \includegraphics[width=.76\textwidth]{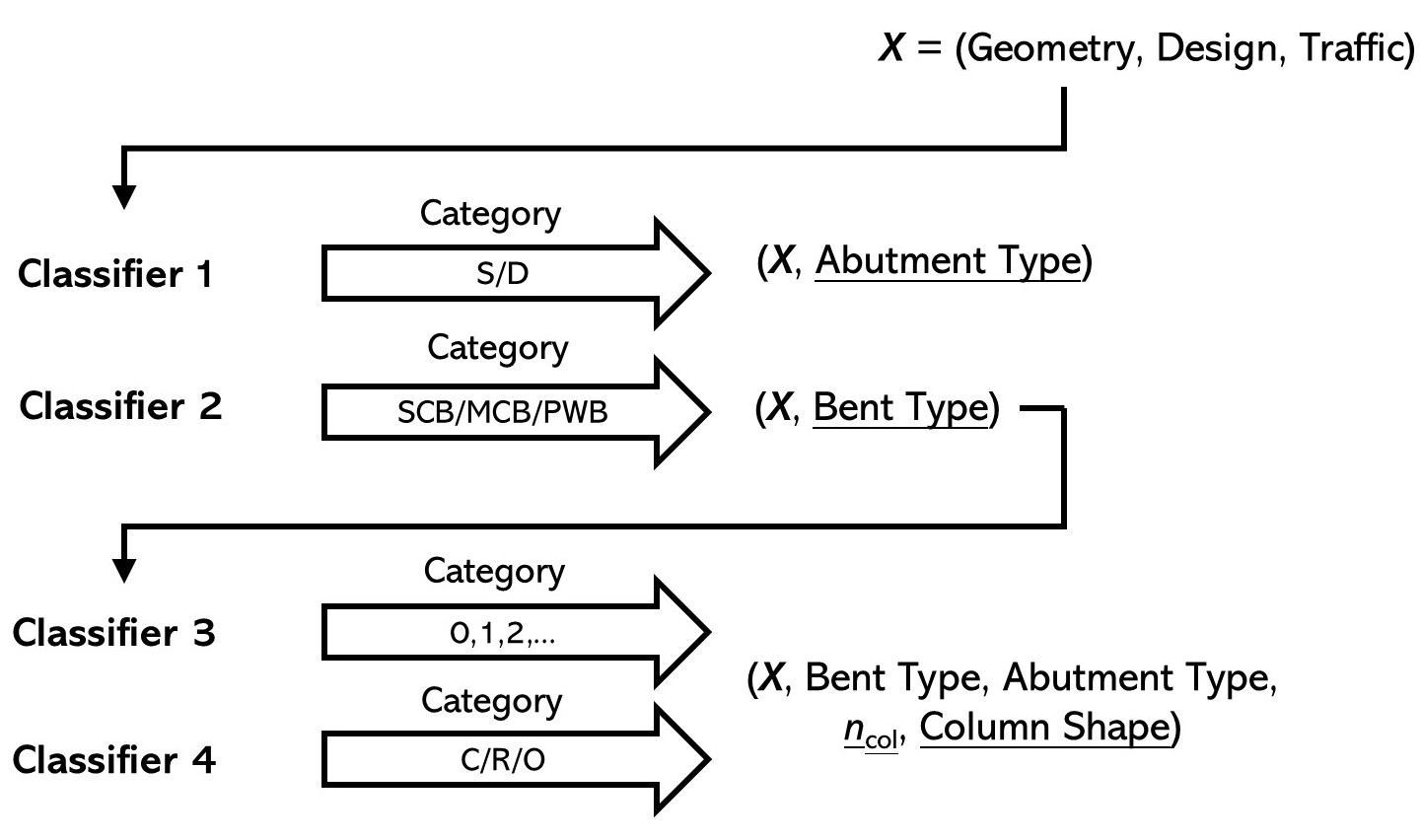}%
	\caption{Proposed classifier chain for imputing missing attributes.}
	\label{fig:chained_classfier}
\end{figure*}

\autoref{fig:chained_classfier} illustrates the ML model architecture used for attribute imputation. Unlike conventional approaches that treat each target attribute independently, this study adopts a classifier chain structure. In this setting, outputs from earlier stages are used as input features for subsequent stages \citep{read2009classifier}. In the first stage, the abutment type and bent type are independently predicted using an ML-based classifier. The predicted bent type is then concatenated with the original input features and passed to the second stage, which predicts the number of columns per bent and column shape. The motivation for using the classifier chain is twofold. First, the chain structure ensures that the engineering logic is maintained among the target attributes. For example, a single-column bent bridge has only one column per bent, while a pier-wall bridge contains one wall per bent. As such, the number of columns becomes meaningful only for multi-column bent bridges. Second, the classifier chain achieved higher predictive accuracy than models that predicted each attribute independently.

\begin{table}[H]
\centering
\captionsetup{justification=centering,singlelinecheck=false}
\caption{Bridge attributes obtained from ML-enabled imputation and their corresponding fields (modified after \cite{chen2025second}.}
\label{tab:bridge_attributes}
\renewcommand{\arraystretch}{1.2}
\begin{tabular}{l|l|l}
\hline
Attribute                  & Valid Fields        & Abbreviation \\ \hline
Bent Type                  & Single-column bent & SCB          \\
                           & Multi-column bent  & MCB          \\
                           & Pier-wall bent     & PWB          \\ \hline
Abutment Type              & Diaphragm-type     & D            \\
                           & Seat-type          & S            \\ \hline
Column Shape               & Circular           & C            \\
                           & Rectangular        & R            \\
                           & Oblong             & O            \\ \hline
Number of columns per bent ($n_{col}$) & 0,1,2,...,7        & --           \\ \hline
\end{tabular}

\raggedright
a: Single-span bridges do not have a bent and are not included in the ML-based predictions.

b: There is a small proportion of bridges that have more than 7 columns per bent. These bridges are considered as outliers and have been excluded before the model training. 
\end{table}

An XGBoost classifier \citep{chen2016xgboost} is used as the predictive model to impute the four target attributes. The hyperparameters of each classifier are tuned using Bayesian optimization \citep{wu2019hyperparameter}, which is conducted within a stratified five-fold cross-validation framework. Early stopping is employed during the training by monitoring validation performance within each fold, such that the boosting process is terminated when no further improvement is observed after a specified number of iterations. The XGBoost was chosen in part because it is well-suited for structured/tabular datasets of limited to moderate size. Given the sample size available in this study, the XGBoost model performed better than a neural network because the latter typically requires substantially larger datasets. Moreover, the proposed framework is intentionally designed to rely solely on tabular data from the NBI. This avoids the need to acquire a large number of images and ensures consistency with the study’s objective of leveraging existing inspection records. However, the entire framework is amenable to computer vision-based classifiers, which typically produce class-specific probability distributions via a softmax layer. 

\begin{figure*}[h]   
 \centering
  \begin{subfigure}[h]{.4\textwidth}
    \hspace{-.15\textwidth} 
	\centering
    \raggedright
	\includegraphics[width=\textwidth]{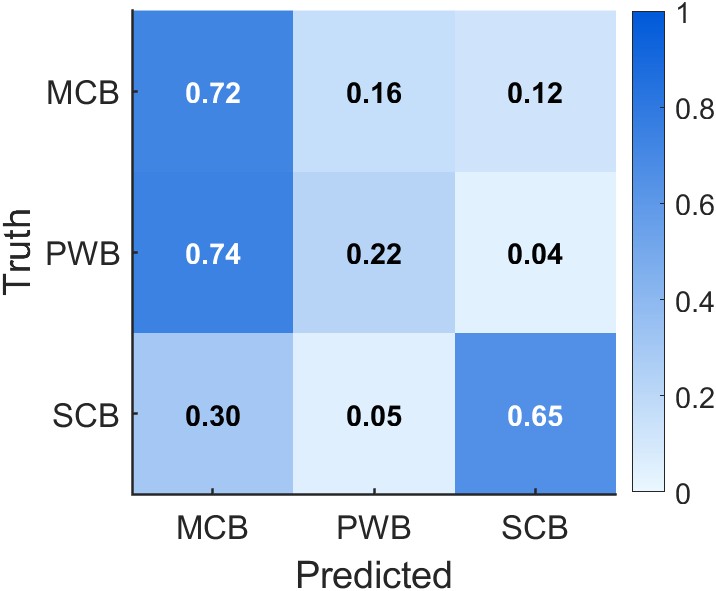}
 	\caption{}
    \label{fig:confusion_benttype}
  \end{subfigure}
  \hspace{0.02\textwidth} 
  \begin{subfigure}[h]{.35\textwidth}
 	\centering
 	\includegraphics[width=\textwidth]{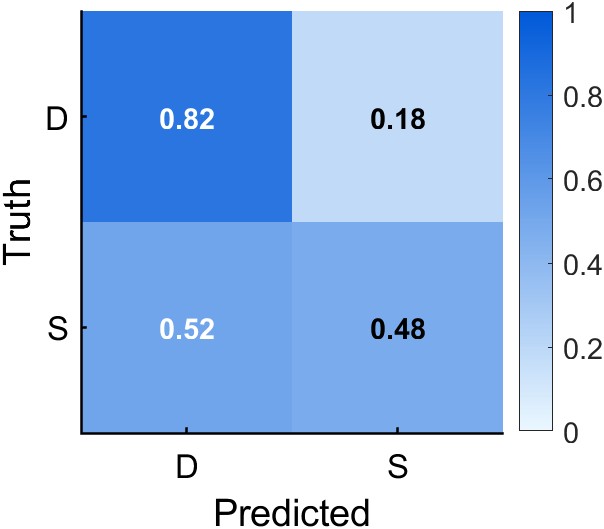}
 	\caption{}
    \label{fig:confusion_abuttype}
  \end{subfigure}
  \begin{subfigure}[h]{.4\textwidth}
 	\centering
 	\includegraphics[width=\textwidth]{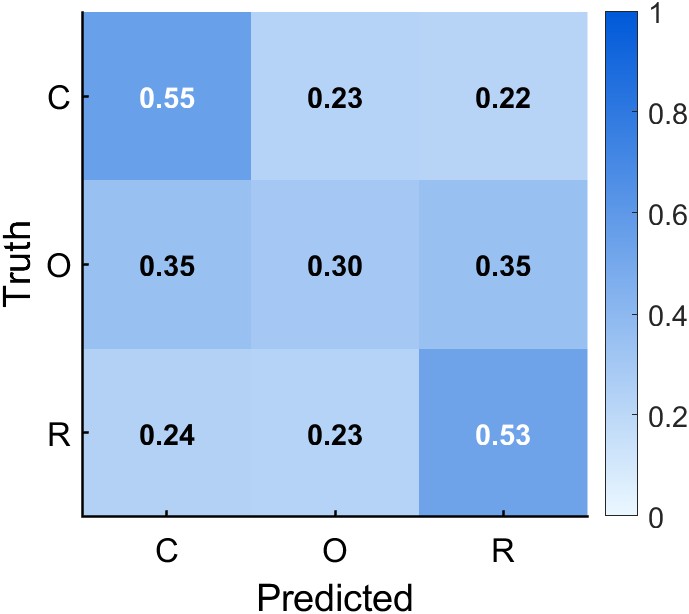}
 	\caption{}
    \label{fig:confusion_colshape}
  \end{subfigure}
  \hspace{0.02\textwidth} 
  \begin{subfigure}[h]{.4\textwidth}
 	\centering
 	\includegraphics[width=\textwidth]{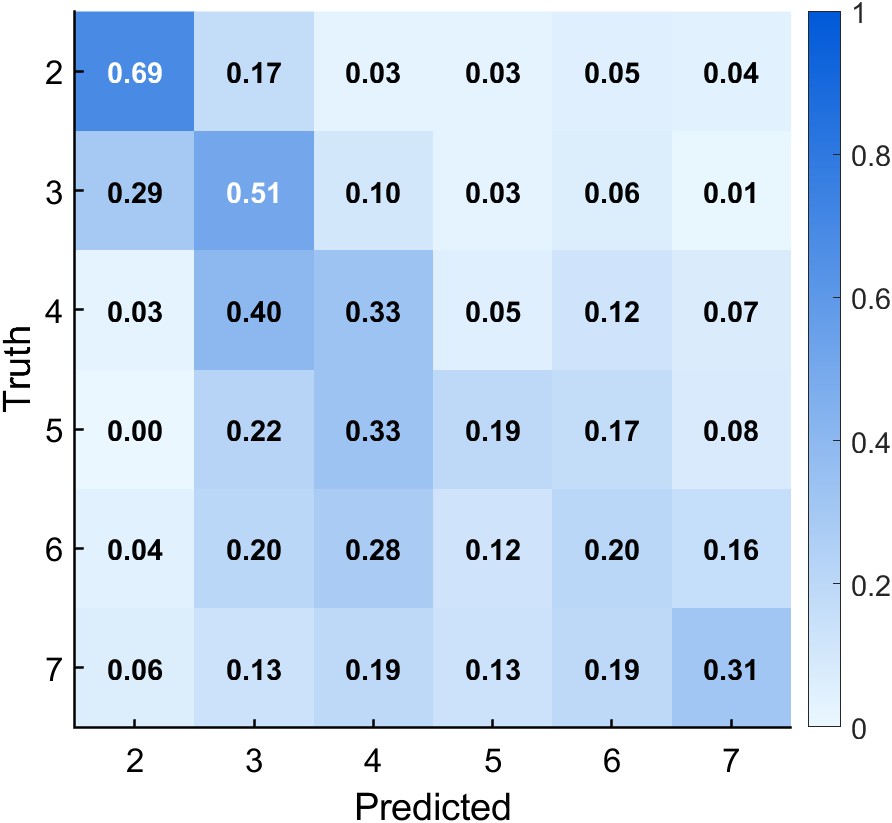}
 	\caption{}
    \label{fig:confusion_numcol}
  \end{subfigure}
 
\caption{Confusion matrices for bridge attribute predictions: (a) bent type. (b) abutment type. (c) column shape. (d) number of columns per bent. Note that (c) applies to single-column and multi-column bent bridges, and (d) only applies to multi-column bent bridges (MCB = Multi-Column Bent, PWB = Pier-Wall Bent, SCB = Single-Column Bent, D=Diaphragm, S=Seat, C = Circular, O = Oblong, and R=Rectangular)}
\label{figs:confusion}
\end{figure*}

The model performance for predicting the four attributes is shown in \autoref{fig:confusion_numcol}. The overall accuracy ranges from 0.40 to 0.68, depending on the specific bridge attribute. The confusion matrices indicate several systematic misclassification patterns. For example, pier-wall bents are often misclassified as multi-column bents, particularly in wider bridges. For the column shape prediction, oblong cross-sectional columns are often confused with circular columns. Additionally, the model performs poorly when predicting bents with 5 and 6 columns.
While achieving high predictive accuracy is not the focus of this study, further improvements can be made by incorporating additional features such as information about the surrounding soil and geotechnical conditions. However, obtaining these attributes would likely require the review of engineering drawings or direct communication with transportation agencies.
The predictive accuracy can also be improved by merging similar attribute categories (e.g., combining oblong and circular sections). However, while this may improve classification performance, it could also reduce accuracy in the subsequent damage and loss assessments, since these properties are known to influence structural behavior. 

\section{Application to Bridge-Specific Risk Assessment} \label{sec:singlebridgescale}

This section investigates how incomplete exposure information affects damage and loss estimation for an individual bridge. To this end, uncertainty quantification is performed for a bridge selected from the NBI. A two-span, multi-column, diaphragm-type, box-girder concrete bridge located in the City of LA (latitude 34.27$^\circ$, longitude -118.52$^\circ$) is considered. The spectral acceleration at a 1-second period ($Sa_{1s}$) is adopted as the intensity measure (IM) in the probabilistic seismic hazard analysis (PSHA) and the subsequent damage and loss assessment. Based on the Caltrans Seismic Design Criteria v2.0 \citep{caltranssdc2018}, and accounting for basin and near-fault effects, the design spectral acceleration is calculated as 0.45 g.

\begin{figure*}[h] 
        \centering
     \includegraphics[width=.95\textwidth]{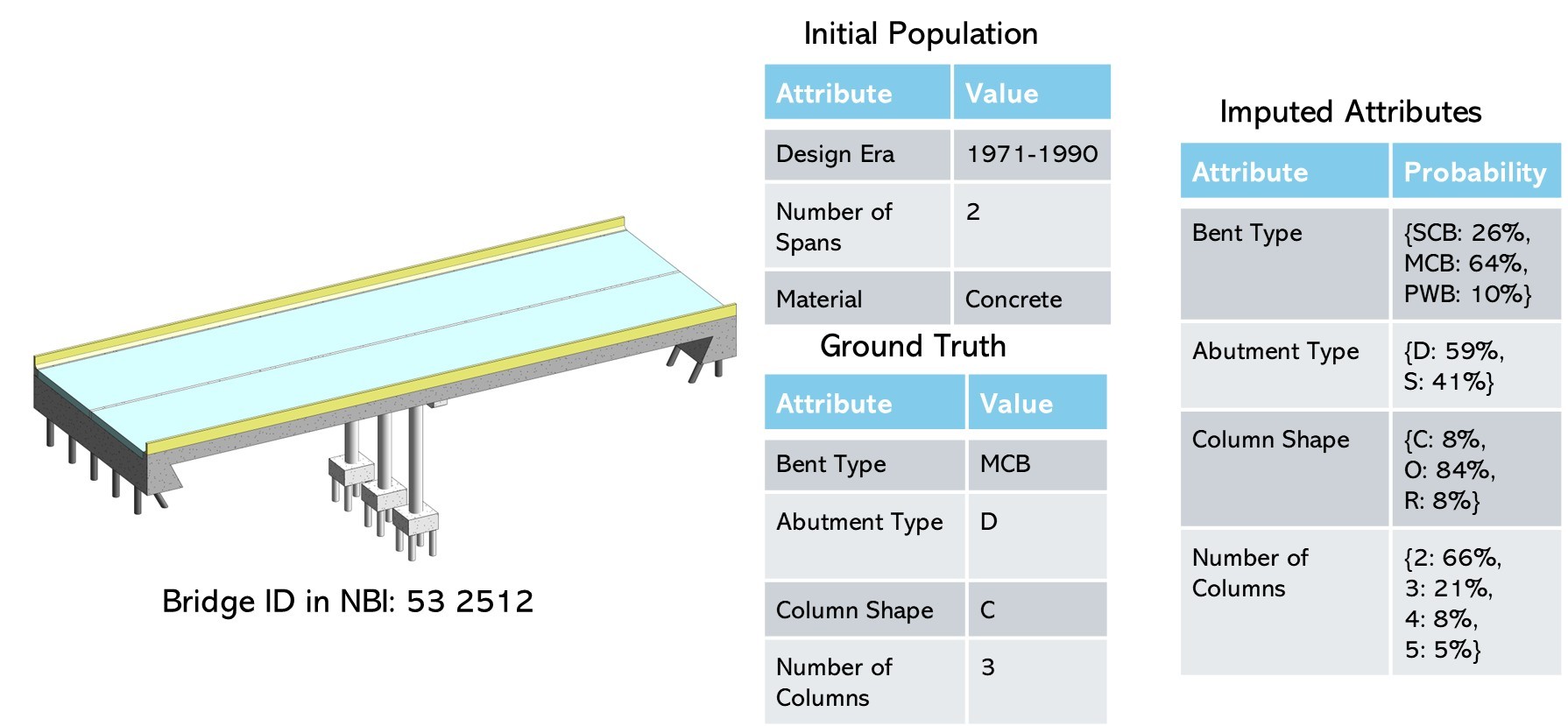}%
	\caption{Schematic illustration of the case study bridge and the relevant exposure information.}
	\label{fig:SingleBridgeTable}
\end{figure*}

Because of the limited information available from the NBI, only three key bridge attributes are directly available: design era, number of spans, and material type. Using these known attributes, together with other related features from the NBI, \autoref{fig:SingleBridgeTable} shows the ML-based prediction of the four missing attributes in terms of their class-specific probability distributions (using \autoref{eq:softmax_calibrated}). The ground-truth values for these originally missing attributes are then obtained from a virtual inspection (also shown in \autoref{fig:SingleBridgeTable}). These ground-truth values provide a basis for evaluating the bias and additional uncertainty in the loss estimates that is introduced by the imputation process.

\begin{figure*}[h]   
 \centering
  \begin{subfigure}[h]{.43\textwidth}
	\centering
	\includegraphics[width=\textwidth]{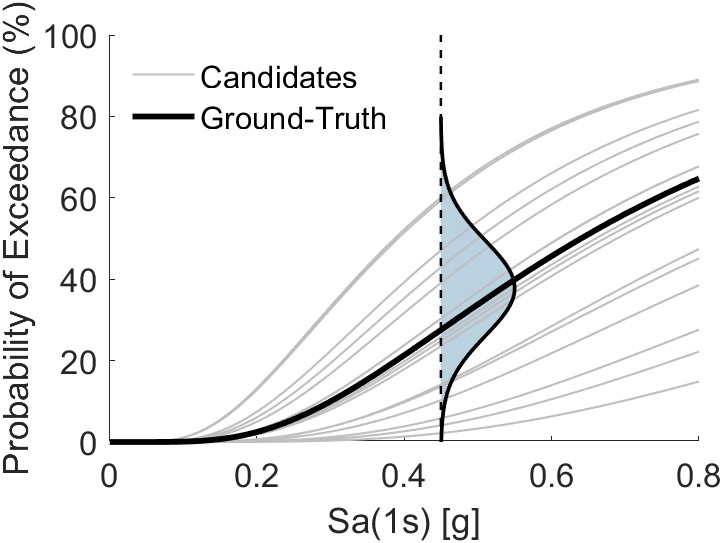}
 	\caption{}
    \label{fig:column_candidate_frag_moderate}
\end{subfigure}
\begin{subfigure}[h]{.55\textwidth}
 	\centering
 	\includegraphics[width=\textwidth]{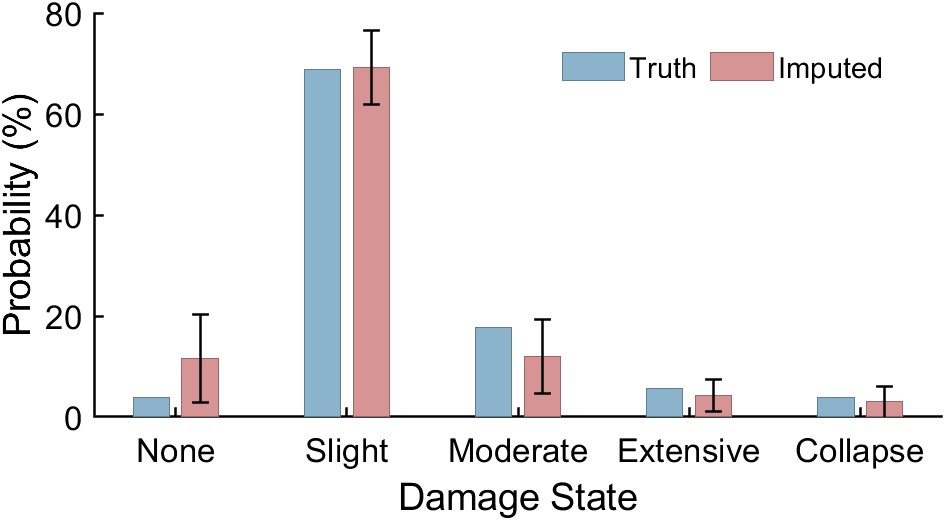}
 	\caption{}
    \label{fig:bar_coldamage_comparison_truth_imputed}
 \end{subfigure}
\caption{(a) Candidate column fragility functions for the moderate damage state. (b) Comparison of column damage probability estimated based on the true and imputed bridge attributes.}
\label{figs:damage_singlebridge}
\end{figure*}

Two evaluation tasks are conducted sequentially to examine how bias and exposure information uncertainty propagate through the damage assessment and loss estimation. In the damage assessment stage, rather than explicitly performing nonlinear response history analyses, component damage states are directly sampled from existing fragility functions. The component fragility database compiled by \cite{chen2025second} is used for this purpose. It classifies bridges into 26 categories based on attributes including design era, span class, bent type, abutment type, and column shape. Because these attributes are predicted with uncertainty (\autoref{fig:SingleBridgeTable}), their combinations yield 16 plausible bridge classes. Consequently, up to 16 candidate fragility functions may be used to assess damage for a given component. For illustration, \autoref{fig:column_candidate_frag_moderate} presents a set of candidate column fragility functions. As shown in \autoref{fig:column_candidate_frag_moderate}, at the specified IM level, this class ambiguity leads to a range of plausible damage probabilities. In contrast, when the ground-truth attributes are known, the bridge class is uniquely defined, resulting in a single deterministic damage probability. \autoref{fig:bar_coldamage_comparison_truth_imputed} presents the estimated damage probabilities for the candidate (i.e., based on imputed-information) and ground-truth bridge classes. The results indicate that bias is introduced due to differences between the imputed and ground-truth damage probability estimates (\autoref{eq:bias_in_state_prob}). Specifically, the estimates based on imputed information exhibit an upward bias in the probability of no damage and slight damage, and downward bias in the probabilities of more severe damage. The uncertainty in the imputed attributes also introduces additional variability in the damage probability estimates, as reflected by the error bars around the mean probabilities (\autoref{eq:epis_in_state_prob}). This variability is not present under the ground-truth class, for which only a single fragility function is applicable. 

\begin{figure*}[h]   
 \centering
  \begin{subfigure}[h]{.49\textwidth}
	\centering
	\includegraphics[width=\textwidth]{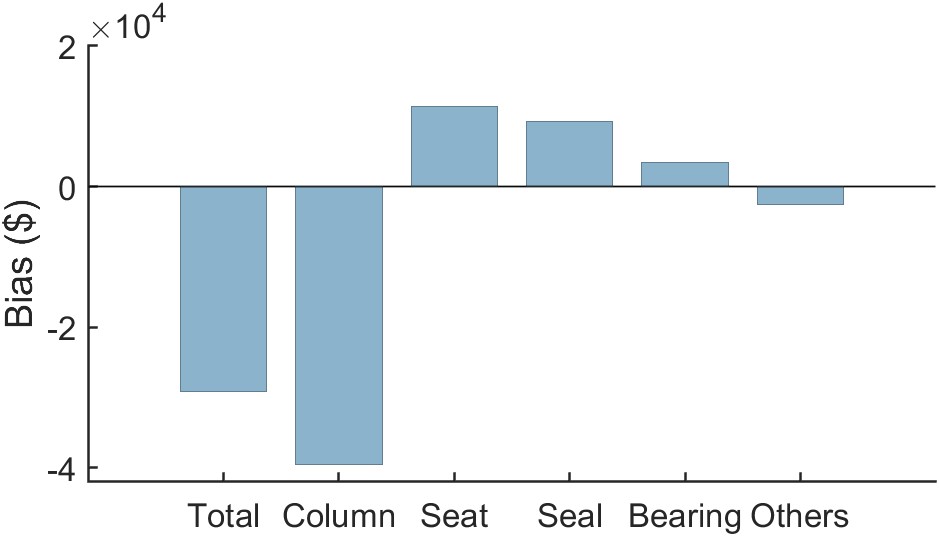}
 	\caption{}
    \label{fig:singlrbridge_loss_bias}
\end{subfigure}
\begin{subfigure}[h]{.5\textwidth}
 	\centering
 	\includegraphics[width=\textwidth]{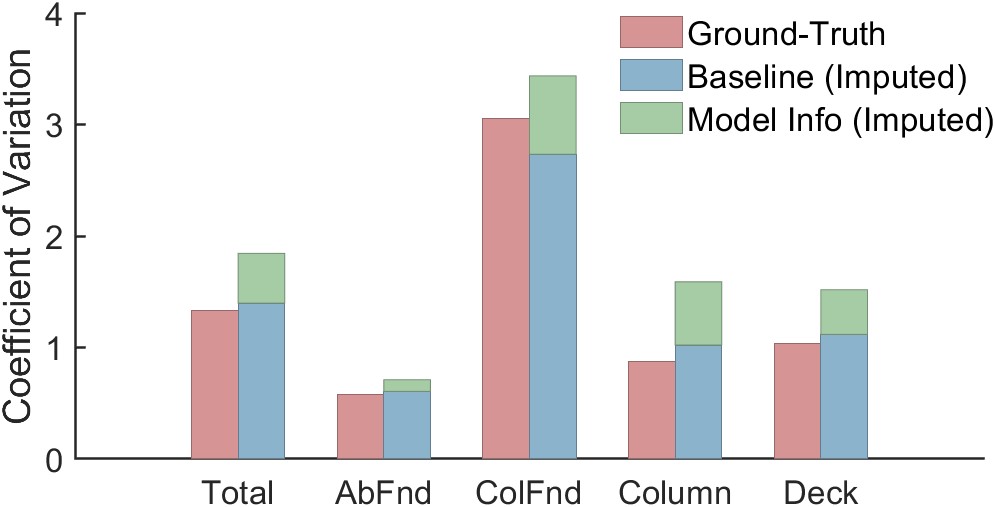}
 	\caption{}
    \label{fig:singlrbridge_loss_CV}
 \end{subfigure}
\caption{(a) Bias and (b) additional uncertainty in the repair cost estimation (AbFnd = Abutment foundation, ColFnd = Column Foundation).}
\label{figs:singlrbridge_loss}
\end{figure*}

The estimated damage states are subsequently used to quantify the bias and additional uncertainty in the loss estimation, following the sampling scheme described in Algorithm \autoref{alg:mc_model_info}. The methodology established in \cite{chen2026high} is adopted, in which the repair cost of a bridge system is decomposed into component-specific repair tasks (e.g., patching and epoxy injection). The corresponding repair costs and associated probability distributions are also provided.
\autoref{fig:singlrbridge_loss_bias} shows the resulting bias in total repair cost, as well as its breakdown across the individual bridge components.
The results show that the estimated repair costs for abutment seat, joint seal, and bearing show an upward bias, 
In contrast, the total repair costs, along with the costs for columns and the sum of other components, exhibit a downward bias. This pattern arises because the model's high-confidence misclassification of the bridge as having seat-type abutments. As a result, components such as the joint seal, bearing, and abutment seat, which are specific to seat-type abutment bridges, are incorrectly included in the loss estimation. \autoref{fig:singlrbridge_loss_CV} presents the decomposition of the coefficient of variation (CV) into the baseline and exposure information uncertainties for the total repair cost, along with its component-wise breakdown, following \autoref{eq:sample:baseline_var} and \autoref{eq:sample:model_info_var}, respectively. The results indicate that, relative to the CV estimated using the true attributes, the CV of the total repair cost estimate increases by 35\% when imputed attributes are used, with exposure information uncertainty contributing 22\% of the total CV. At the component level, the increase in CV ranges from 18\% to 80\%, depending on the component considered.

\section{Application to City-Scale Assessment of Bridge Networks} \label{sec:cityscale}
\begin{figure*}[h]   
 \centering
  \begin{subfigure}[h]{.46\textwidth}
	\centering
	\includegraphics[width=\textwidth]{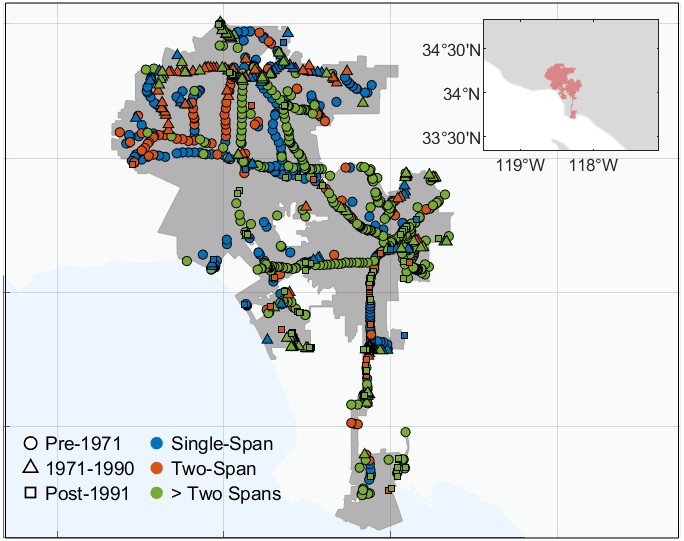}
 	\caption{}
    \label{fig:distribution_bridges}
\end{subfigure}
\begin{subfigure}[h]{.5\textwidth}
 	\centering
 	\includegraphics[width=\textwidth]{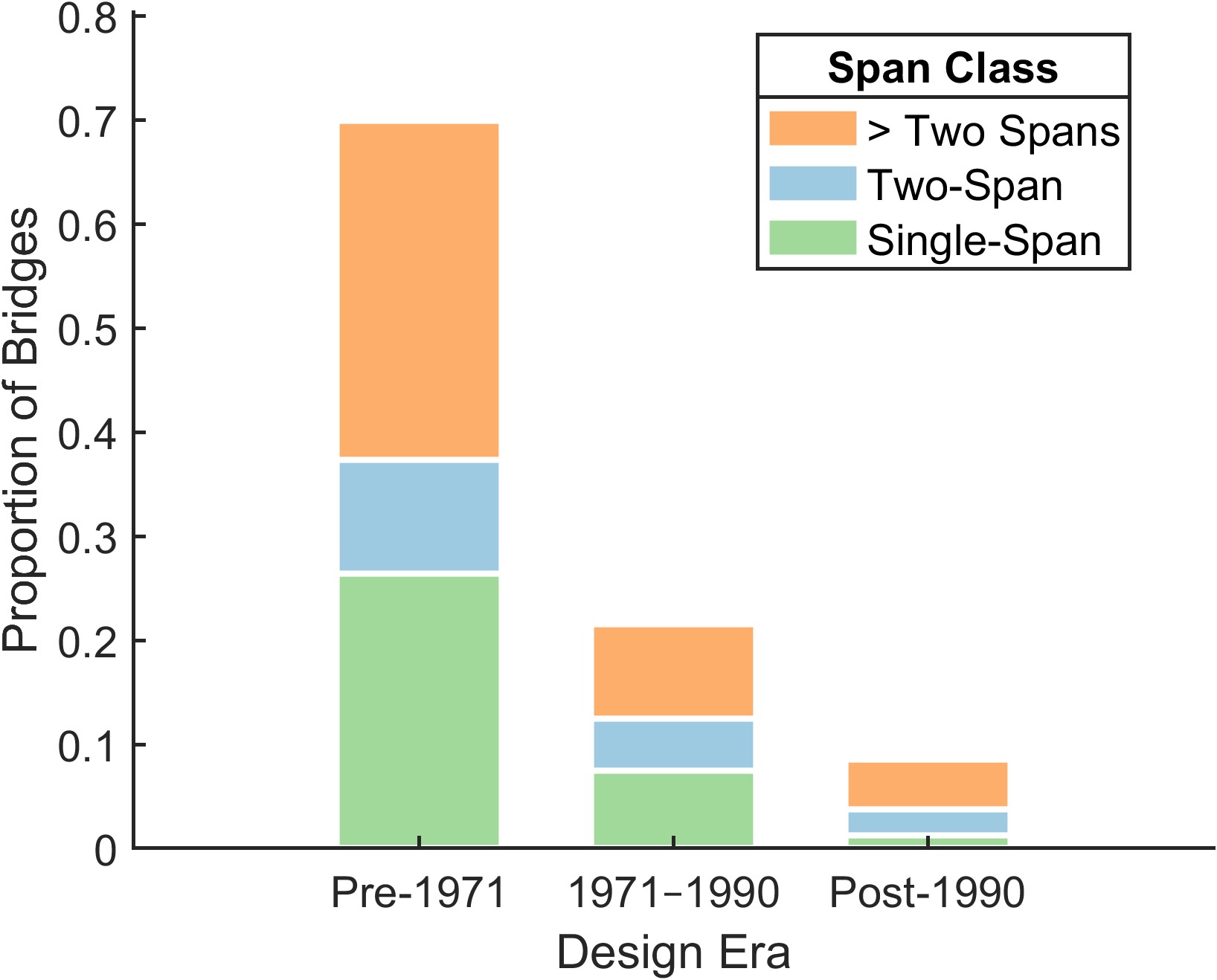}
 	\caption{}
    \label{fig:Bar_BridgeAttributes}
 \end{subfigure}
\caption{(a) Spatial distribution of the City of LA bridge network. (b) Distribution of the bridge design era and span class from the initial population.}
\label{figs:bridgeassets}
\end{figure*}

This section evaluates the impact of incomplete exposure information on regional economic loss estimation. We use the 1,179 bridges in the City of LA network as the testbed, which is shown in \autoref{fig:distribution_bridges}. Following the bridge exposure inventory augmentation framework shown in \autoref{fig:framework_augmentation}, the year built and span type are first retrieved from the NBI, and their corresponding distributions are shown in \autoref{fig:Bar_BridgeAttributes}. It shows that the network is dominated by single-span and multi-span ($>$ 2 spans) bridges constructed before 1971 i.e., with minimal seismic design considerations.

Recall in \autoref{sec:modeltrainingevaluation} that the City of LA network served as the test set for evaluating the performance of the ML models trained on statewide representative bridges (\autoref{figs:confusion}). The developed models predict probability distributions for each missing attribute required for component-level loss estimation (\autoref{tab:bridge_attributes}). This section further examines how such probabilistic imputations influence the repair cost estimation at the regional scale.

\subsection{Hazard characterization and fragility database}
Regional hazard characterization is performed to evaluate the joint distribution of shaking intensities at the individual bridge sites for a given rupture scenario. The $\bm{\textbf{M}}_w$ 7.66 rupture scenario from \cite{Wu_etal_regional_recovery} is adopted based on site-specific hazard disaggregation and the UCERF3 seismic source model.  The corresponding ground motion random field (GMRF) of $\boldsymbol{IM}$ for a given scenario, $rup$, is represented through a joint lognormal distribution \citep{baker2021seismic}:

\begin{equation} \label{eq:jointlogIM}
	\ln (\boldsymbol{IM} \mid rup)  \sim \mathcal{N}( \boldsymbol{\mu}(\ln \boldsymbol{IM}\mid rup), \boldsymbol{\Sigma}(\ln \boldsymbol{IM} \mid  rup)), 
\end{equation}

where $\boldsymbol{\mu}(\ln \boldsymbol{IM}|rup)$ is the mean vector corresponding to $\ln IM$ for each bridge site, and $\boldsymbol{\Sigma}(\ln \boldsymbol{IM} \mid rup)$ is the covariance matrix. The latter can be written as:

\begin{equation} \label{eq:covmat}
	\boldsymbol{\Sigma}(\ln (\boldsymbol{IM}|rup)) = \tau^2\boldsymbol{1} + \mathrm{diag}(\pmb{\phi}) \;\boldsymbol{R} \;\mathrm{diag}(\pmb{\phi}), 
\end{equation}

where $\tau$ is the between-event standard deviation (constant for a given scenario), $\pmb{\phi}$ is a vector of site-specific within-event standard deviations, $\boldsymbol{1}$ is an identify matrix, $\mathrm{diag}(\cdot)$ represents a diagonal matrix, and $\boldsymbol{R}$ is the within-event correlation matrix with ones on the diagonal and correlation of the within-event residuals between sites $j$ and $j^\prime$ ($\rho_{j,j\prime}$) are the off-diagonal elements.

\begin{figure*}[h] 
        \centering
     \includegraphics[width=.49\textwidth]{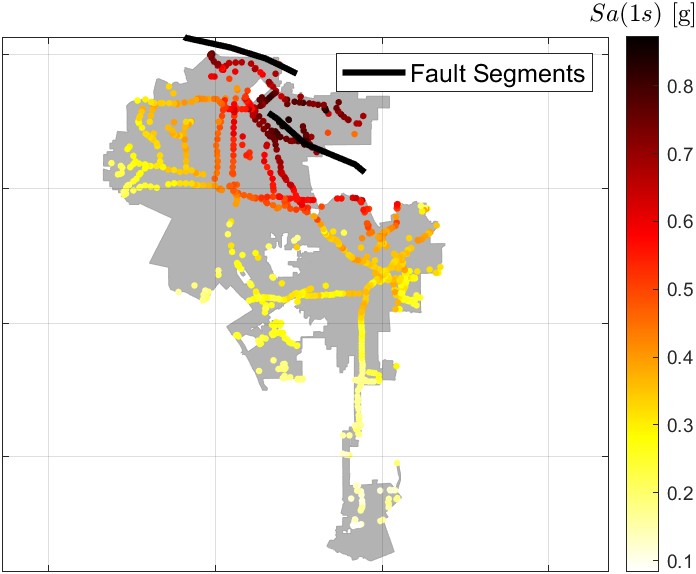}%
	\caption{Median GMRF and causative fault segments for the considered rupture scenario.}
	\label{fig:medianGMRF}
\end{figure*}

One hundred ground motion maps are sampled from this joint distribution using the \cite{boore2014nga} and \cite{jayaram2009correlation} ground motion and spatial correlation models, respectively. 
\autoref{fig:medianGMRF} shows the median $Sa(1s)$ at each bridge site, and the causative fault segments corresponding to the selected scenario. Multiple fault segments contribute to this scenario because the UCERF3 framework allows rupture propagation across different faults to satisfy a set of geophysical criteria \citep{field2014uniform}.

\begin{figure*}[h] 
        \centering
     \includegraphics[width=.75\textwidth]{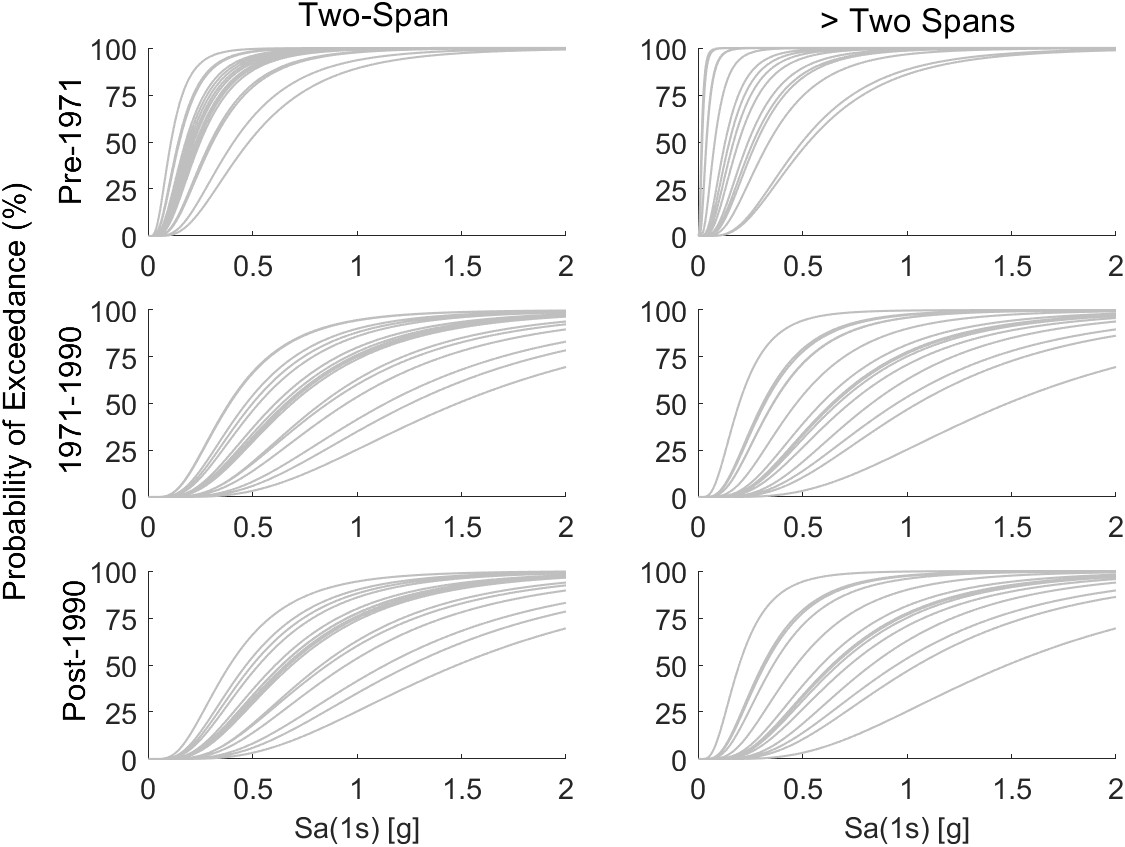}%
	\caption{Candidate column fragility functions for the moderate damage state conditioned on the design era and span class.}
	\label{fig:regional_candid_fragilities}
\end{figure*}

Similar to \autoref{sec:singlebridgescale}, damage states are sampled from component-level fragility functions compiled by \cite{chen2025second}. However, due to uncertainties in the predicted bridge attributes, the assigned fragility functions for each bridge are no longer deterministic but instead exhibit variability. \autoref{fig:regional_candid_fragilities} shows sets of candidate column fragility functions conditioned on the two NBI attributes. The results indicate that a wide range of damage probabilities, rather than a single estimate, may be associated with a given bridge. This highlights the additional uncertainty introduced into the subsequent loss assessment.

\subsection{Bias and additional uncertainty in regional economic loss estimation}

The repair cost for each bridge is sampled for each simulated ground motion map. These bridge-specific samples are then aggregated across the inventory and all ground motion maps to obtain samples of regional repair costs. For each bridge and ground motion map, the damage to its constituent components is first sampled using the available fragility functions. Based on the resulting component-level damage states, the corresponding repair tasks are identified. In cases of irreparable damage triggered by either complete column failure or span unseating, the repair task is a full bridge replacement. Once the repair tasks are determined, the required material quantities can be estimated. Repair costs are then sampled based on the material quantities and their associated unit costs. A detailed description of the repair cost estimation procedure is provided in \cite{chen2026high}. When imputed attributes are used, an additional step is introduced prior to damage sampling. Specifically, a bridge class is first sampled according to the probability distributions of the imputed attributes. A corresponding set of fragility functions associated with the sampled bridge class is then assigned for that realization. Based on the assigned bridge class and the associated fragilities, the subsequent damage and repair cost sampling procedures are conducted.

\begin{figure*}[h]   
 \centering
  \begin{subfigure}[h]{.32\textwidth}
	\centering
	\includegraphics[width=\textwidth]{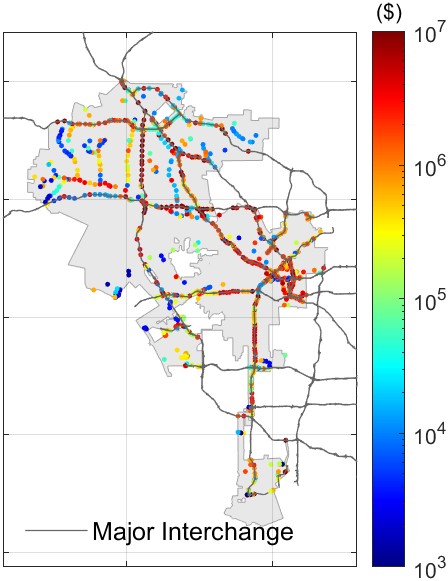}
 	\caption{}
    \label{fig:Regional_MeanLossTrueMap}
\end{subfigure}
\begin{subfigure}[h]{.32\textwidth}
 	\centering
 	\includegraphics[width=\textwidth]{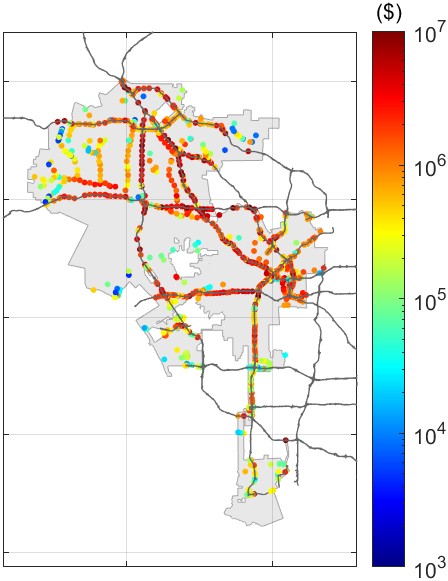}
 	\caption{}
    \label{fig:Regional_MeanLossImputedMap}
 \end{subfigure}
 \begin{subfigure}[h]{.32\textwidth}
 	\centering
 	\includegraphics[width=\textwidth]{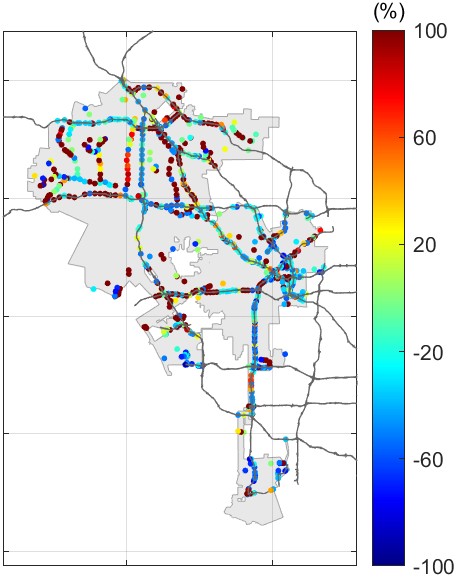}
 	\caption{}
    \label{fig:Regional_Bias_pctDiff}
 \end{subfigure}
\caption{Spatial distribution of mean economic loss estimated using (a) true and (b) imputed bridge attributes. (c) Bias caused by the imputed estimates relative to the ground truth}
\label{figs:Mean_Regional_Loss}
\end{figure*}

\autoref{figs:Mean_Regional_Loss} shows the spatial distribution of bridge-specific mean losses. The overall spatial pattern of losses is preserved when imputed attributes are used, with the highest expected losses concentrated along primary corridors and major interchanges in the network. However, notable discrepancies are observed in \autoref{fig:Regional_Bias_pctDiff}, with the bias ranging from -100\% to +100\%. Bridges located along major highway interchanges tend to exhibit systematic downward bias, in some cases approaching 100\%. This suggests that attribute imputation may preferentially assign less vulnerable bridge classes in these high-consequence regions, thereby underestimating the losses. 

\begin{figure*}[h]   
 \centering
  \begin{subfigure}[h]{.31\textwidth}
	\centering
	\includegraphics[width=\textwidth]{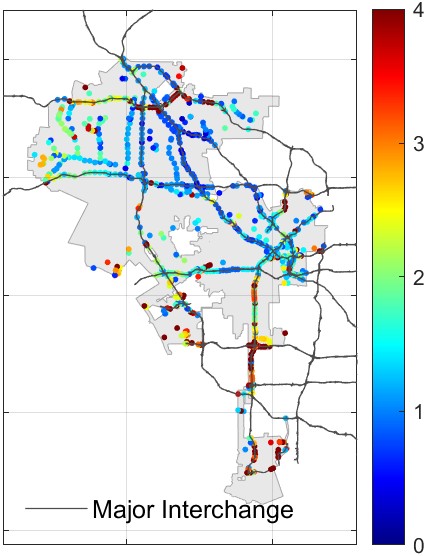}
 	\caption{}
    \label{fig:Regional_CV_baseline}
\end{subfigure}
\begin{subfigure}[h]{.32\textwidth}
 	\centering
 	\includegraphics[width=\textwidth]{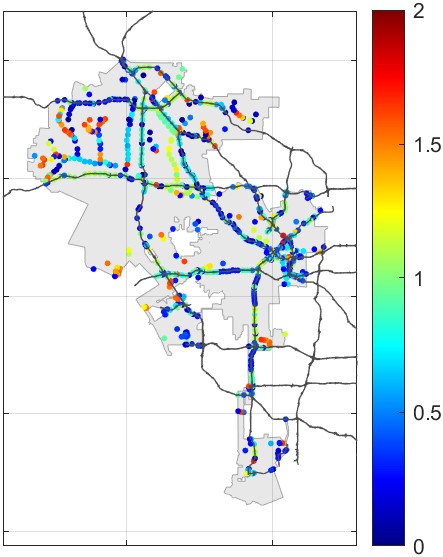}
 	\caption{}
    \label{fig:Regional_CV_model}
 \end{subfigure}
 \begin{subfigure}[h]{.33\textwidth}
 	\centering
 	\includegraphics[width=\textwidth]{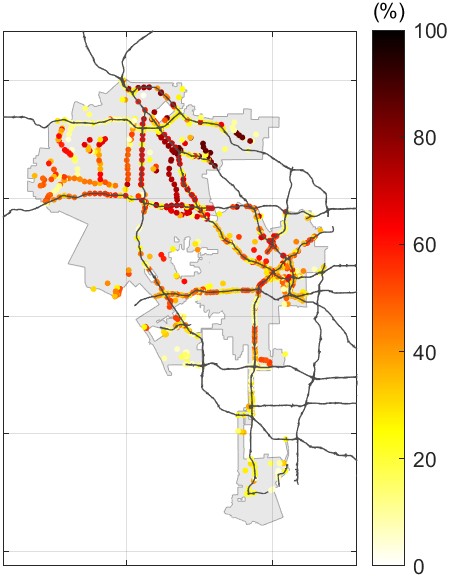}
 	\caption{}
    \label{fig:Regional_pct_model_CV}
 \end{subfigure}
\caption{Spatial distribution of the coefficient of variation contributed by (a) the baseline and (b) exposure information uncertainty. (c) Relative contribution of exposure information uncertainty to the total coefficient of variation.}
\label{figs:Regional_CV_Model_Base}
\end{figure*}

The bridge-specific CV is decomposed into the baseline and exposure information components (\autoref{fig:Regional_CV_baseline} and \autoref{fig:Regional_CV_model}), with the relative contribution of the latter shown in \autoref{fig:Regional_pct_model_CV}. The results indicate that the baseline uncertainty remains dominant across much of the network, suggesting that uncertainties in the hazard characterization, damage assessment, and repair cost estimation, dominate the overall dispersion in bridge-specific loss estimates. In comparison, the CV attributable to exposure information is generally smaller in magnitude, accounting for less than 40\% of the total CV for most bridges. Notably, however, many bridges located at major highway interchanges exhibit exposure information contributions exceeding 50\% of the total CV. This pattern highlights the value of targeted inspections for these functionally critical bridges, where reducing class ambiguity could meaningfully improve loss estimation. Such a strategy is more effective than attempting to improve inventory completeness uniformly across the entire network. This issue is further examined in the next subsection. 

The bridge-specific repair cost samples are aggregated to estimate regional economic loss. Their variance is further decomposed into the baseline and exposure information components following \autoref{eq:RL_var_decompose} and the procedure described in the Appendix. \autoref{figs:compare_mean_CV} summarizes the bias and CV of the regional loss estimates. Compared to when the true bridge attributes are used, which yields a mean loss estimate of approximately 2.12 billion dollars, the imputation of missing attributes underestimates the mean regional loss by 12.8\%. Similarly, while the use of true bridge attributes results in a CV of 0.32, this value increases by 11.1\% when imputed attributes are adopted. The result indicates that the baseline component accounts for the majority of the total CV because the total uncertainty is dominated by sources other than incomplete exposure information. The relative contribution of each uncertainty source is further examined in a later subsection on sensitivity analysis.

\begin{figure*}[h]   
 \centering
  \begin{subfigure}[h]{.47\textwidth}
	\centering
	\includegraphics[width=\textwidth]{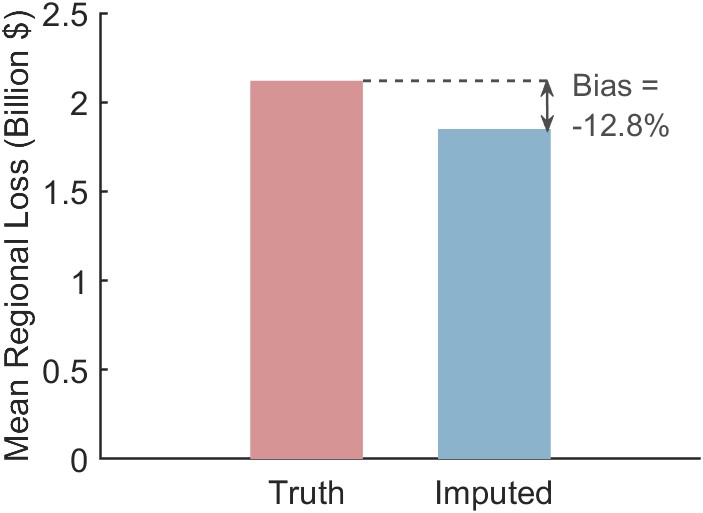}
 	\caption{}
    \label{fig:Compare_Total_Mean_Loss}
\end{subfigure}
\begin{subfigure}[h]{.52\textwidth}
 	\centering
 	\includegraphics[width=\textwidth]{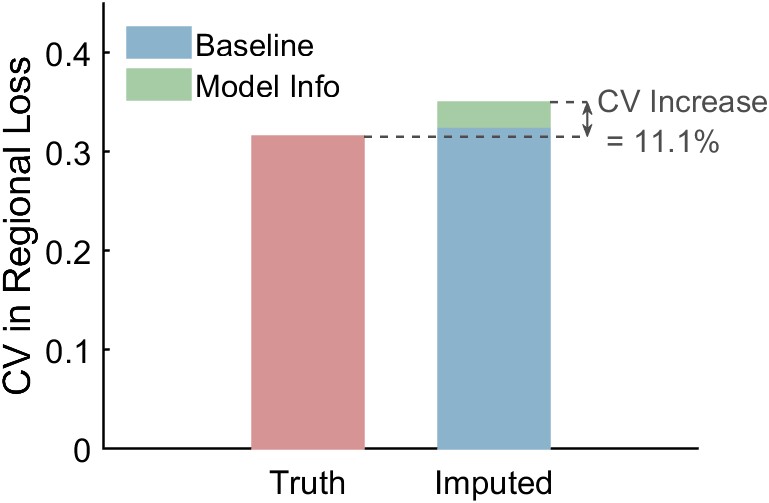}
 	\caption{}
    \label{fig:Compare_CV_Regional_Loss}
 \end{subfigure}
\caption{Comparing the mean and coefficient of variation (CV) of the regional loss estimates when the true and imputed attributes are used: (a) mean; (b) coefficient of variation.}
\label{figs:compare_mean_CV}
\end{figure*}

\subsection{Uncertainty-informed inspection prioritization} \label{sec:inspectionprior}
\autoref{fig:Compare_CV_Regional_Loss} shows that imperfect exposure information increases the variance in regional economic loss estimates, highlighting the importance of targeted inspections to identify true bridge attributes to reduce this source of uncertainty.

\begin{figure*}[h]   
 \centering
  \begin{subfigure}[h]{.45\textwidth}
	\centering
	\includegraphics[width=\textwidth]{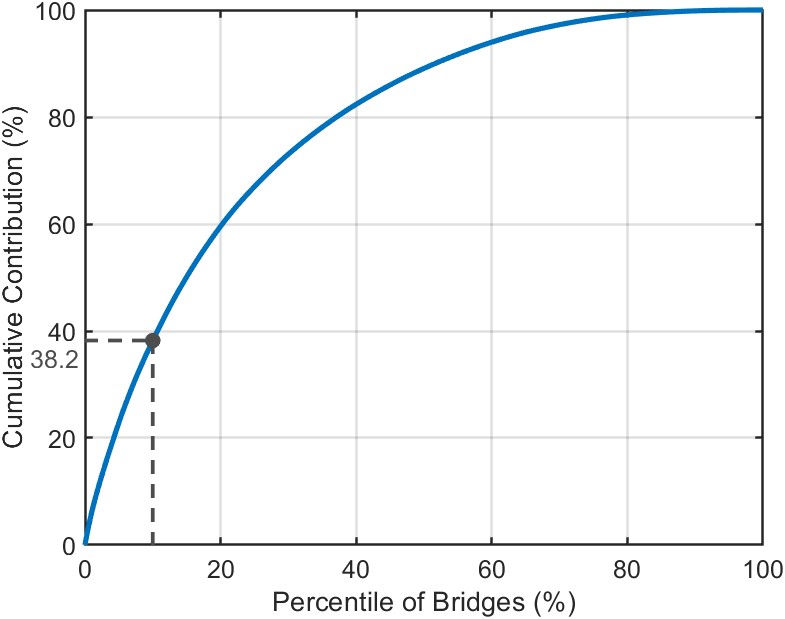}
 	\caption{}
    \label{fig:CumulativeContribution}
\end{subfigure}
\begin{subfigure}[h]{.3\textwidth}
 	\centering
 	\includegraphics[width=\textwidth]{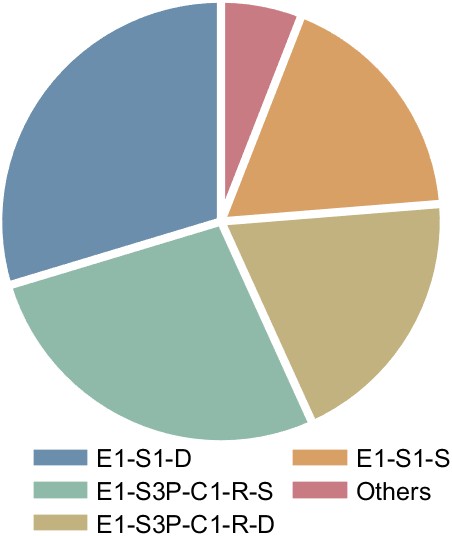}
 	\caption{}
    \label{fig:Pie_BridgeClasses}
 \end{subfigure}
\caption{(a) Cumulative exposure information uncertainty function. (b) Bridge classes associated with the top 10\% of contributors (E1 = Pre-1971. S1 = Single-Span. C1 = Single-Column Bent. R = Rectangular Section. S = Seat-Type Abutment. D = Diaphragm-Type Abutment.).}
\label{figs:inspection}
\end{figure*}

Based on the bridge-specific uncertainty shown in \autoref{fig:Regional_CV_model}, the cumulative distribution of exposure information uncertainty is evaluated by ranking bridges in descending order of their corresponding contributions (\autoref{fig:CumulativeContribution}). The cumulative function has a logarithmic form, with the top 10\% of bridges accounting for 38.2\% of the total exposure information uncertainty. 

\autoref{fig:Pie_BridgeClasses} further disaggregates the top 10\% of contributors by their ground-truth bridge classes. The results indicate these high-contributing bridges are predominately either single-span (S1) or have more than two spans (S3P) and a single column per bent (C1). In addition, they are mostly pre-1971 (E1) bridges, which are inherently vulnerable due to the lack of modern seismic design considerations. This observation can be explained through \autoref{eq:sample:model_info_var}, which shows how different bridge classes contribute to exposure information uncertainty. Specifically, \autoref{eq:sample:model_info_var} indicates that large contributions are associated with bridge classes that have class-specific mean losses that are substantially different from the overall mean (i.e., $\mu_i$ is far from $\mu$) and high probabilities of occurrence (i.e., $\pi_i$ is large). For S1 bridges, the primary source of discrepancy arises from abutment type misclassification. Seat-type bridges are more vulnerable due to the potential for unseating, which can lead to costly repair or even full replacement. In contrast, diaphragm-type bridges are more robust and do not exhibit this failure mode. These two bridge classes represent two extremes in expected repair costs, and the misclassification of abutment type leads to substantial deviation from the class-specific overall mean losses. For S3P-C1 bridges, the main source of uncertainty is bent configuration misclassification. These bridges are incorrectly identified as having multi-column bents, which are associated with higher vulnerability and repair costs. This misclassification systematically overestimates the mean loss for this class. In both cases, systematic misclassification shifts the class-specific mean losses away from the overall mean (i.e., the $\mu_i$ and $\mu$ terms in \autoref{eq:sample:model_info_var}, respectively), thereby increasing their contribution to exposure information uncertainty.

A what-if scenario is used to investigate the effect of uncertainty reduction through targeted inspection. Consider the situation where an additional round of inspection is performed in which the true attributes of these top 10\% of contributors are obtained (corresponding to 118 bridges in the City of LA network). The imputed attributes for these inspected bridges are then replaced with their ground-truth values, while the remaining 90\% of bridges continue to use imputed attributes from the ML model predictions. The regional risk assessment is then performed using the workflow described in the preceding subsection.

\autoref{tab:bias_uncertainty_reduction} summarizes the associated change in bias and uncertainty. The results show that this targeted inspection effectively reduces the CV associated with exposure information uncertainty by 40.3\%. In addition, the bias in mean loss estimation is reduced by 12.2\%. However, the reduction in the total CV is relatively modest (2.9\%), indicating that other sources of uncertainty still dominate the overall variability (as discussed in the next subsection).
Overall, these findings suggest that a substantial portion of exposure information uncertainty can be mitigated by inspecting a small subset of bridges, with the added benefit of a modest reduction in bias.

\begin{table}[H]
\centering
\caption{Summary of bias and uncertainty reduction after inspection.}
\label{tab:bias_uncertainty_reduction}
\begin{tabular}{lc}
\toprule
 & Percentage Reduction (\%) \\
\midrule
Bias & 12.2 \\
Coefficient of Variation (Exposure Information) & 40.3 \\
Coefficient of Variation (Total) & 2.9 \\
\bottomrule
\end{tabular}
\end{table}

\subsection{Disaggregation of uncertainty by risk assessment step}
Previous sections aggregated all sources of uncertainty (i.e., GMRF, damage, exposure information, and loss) into the regional economic loss assessment. This section conducts one-way sensitivity analyses to investigate the relative contribution of each source to the uncertainty in the regional loss estimation. In each run of the sensitivity analysis, only one source of uncertainty is considered while all other variables are considered deterministic. For continuous variables (i.e., GMRF and loss), the values are fixed at the median. For discrete variables (i.e., damage and exposure information), the most probable category is used.

\begin{figure*}[h] 
        \centering
     \includegraphics[width=.55\textwidth]{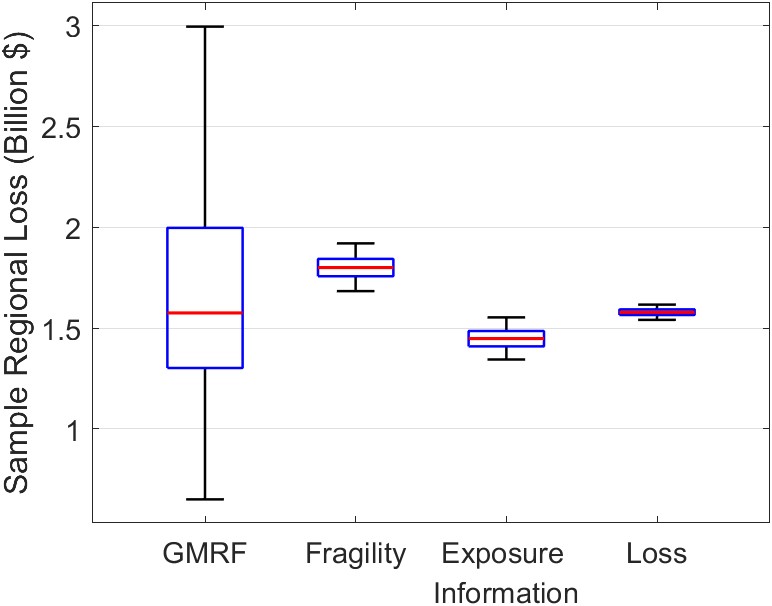}%
	\caption{Sensitivity analysis showing relative contribution of each source of uncertainty to the total variance (GMRF: Ground Motion Random Field).}
	\label{fig:sensitivity}
\end{figure*}

\autoref{fig:sensitivity} shows the results of the sensitivity analyses, where the whiskers indicate the $5^{\text{th}}-95^{\text{th}}$ quantile range of regional economic loss, and the inner boxes represent the $25^{\text{th}}-75^{\text{th}}$ quantile range. The GMRF is shown to be the largest contributor to the uncertainty in seismic risk assessment, which is consistent with prior studies \citep{ching2009propagating,chen2026high}. Conversely, the uncertainty associated with the loss functions has the smallest contribution. The damage fragilities and exposure information have comparable contributions to the overall uncertainty. This observation suggests that the lack of exposure information inflates the total uncertainty to a level equivalent to doubling that of the damage assessment stage. However, this uncertainty can be reduced by observing only a small subset of bridges, as described in \autoref{sec:inspectionprior}.

\section{Conclusions}
This study addresses a gap that is often overlooked in regional risk assessment by focusing on the uncertainty introduced by incomplete exposure data and its probabilistic imputation. While previous work has extensively quantified uncertainties in hazard, fragility, and loss modeling, this study emphasizes the role of uncertainty in exposure modeling. This form of epistemic uncertainty, which is introduced when missing asset attributes are imputed rather than directly observed, is a distinct and non-negligible contributor to the overall risk uncertainty. 

Exposure information uncertainty is quantified and isolated from other sources by leveraging the law of total variance. This provides a clear mathematical separation between the uncertainty that is attributable to incomplete exposure information and the baseline uncertainty stemming from other sources. The proposed methodology integrates analytical formulations with a Monte Carlo-based sampling scheme. In doing so, it tracks how misrepresentation of exposure-based attributes propagates through the damage and loss estimation stages, introducing both bias and additional uncertainty in regional loss estimates. 

To support the uncertainty quantification, a comprehensive framework is developed for generating complete asset inventories. The framework defines required attributes, constructs an initial population from public datasets, and imputes missing information using both predictive models and engineering rulesets. It is then implemented to enable a high-resolution, high-fidelity seismic risk assessment of a highway bridge network in Los Angeles, California. Four key attributes that are required for risk assessment but absent from the NBI are obtained through statewide virtual inspections of over 1,600 California representative bridges. These attributes, which include the bent and abutment types, column shape, and number of columns per bent, are subsequently predicted using machine learning (ML) models. The performance of the ML models is evaluated using the City of Los Angeles (LA) bridge inventory. For each attribute, the developed models generate both point estimates and class-specific probability distributions, serving as the basis for uncertainty quantification in subsequent regional damage and loss estimation.

The proposed methodology and the augmented bridge exposure inventory are applied to both bridge-specific and city-scale risk assessments. For the bridge-specific assessment, component-level damage is sampled from existing fragility functions. Unlike conventional damage assessment in which only a single, deterministic set of fragilities is assigned, multiple candidate fragility functions can be associated with the considered bridge due to ambiguity in the exposure class. This ambiguity introduces additional uncertainty in damage predictions, which in turn propagates into the repair cost estimation, leading to both bias and increased variability.

At the regional scale,the framework is applied to the City of LA network, which comprises more than one thousand bridges. The results show that the uncertainty in exposure information can significantly increase the overall variability in the regional loss estimates and introduce systematic bias. Additionally, the spatial pattern indicates that the exposure information uncertainty tends to concentrate in critical areas, such as major interchanges. While these effects are substantial, this variability can be effectively reduced by inspecting a relatively small subset of the bridges, providing a practical strategy for uncertainty reduction.

Overall, these findings suggest that exposure characterization should not be treated as a deterministic step in regional risk assessment, especially when detailed asset attributes are not available, and probabilistic missing data imputation is adopted. Incorporating and explicitly quantifying exposure information uncertainty is therefore essential for producing more reliable and informed risk estimates.

\begin{appendices}
\section{Sampling Scheme for Regional Economic Loss} \label{app:sampling_scheme}
As described in the main text, the law of total variance decomposes the total variability of regional economic loss ($RL$) into two components: the baseline variance and the exposure information variance. 
The baseline variance ${\mathrm{Var}^{\text{base}}(RL)}$ would remain even if the correct class was known, whereas the exposure information variance ${\mathrm{Var}^{\text{e}}(RL)}$ stems only from imperfect exposure imputation. 

The baseline variance in \autoref{eq:total_var_regional} can be expressed as: 
\begin{equation*}
\begin{split}
    \mathrm{Var}^{\text{base}}(RL) &= \mathbb{E}\!\left[\mathrm{Var}\left(\sum_{b=1}^BL_b | \mathcal{E}\right)\right] \\
    &= \mathbb{E}\!\left[ \sum_{b=1}^B \mathrm{Var}(L_b|\mathcal{E}) + 2\sum_{b<b^\prime}\mathrm{Cov}(L_b,L_{b^\prime}|\mathcal{E})   \right] \\
    &= \sum_{b=1}^B\mathbb{E}\left[\mathrm{Var}(L_b|\mathcal{E}) \right] + 2 \sum_{b<b^\prime}\mathbb{E}\left[\mathrm{Cov}(L_b, L_{b^\prime}|\mathcal{E})\right]
\end{split}
\end{equation*}
This decomposes the baseline variance into two parts. The first part is the sum of the baseline variances of individual bridges, and can be estimated using the same technique introduced in \autoref{sec:uncertainty_loss}. The second part estimates the expected inter-bridge loss covariance, which is introduced as follows. 

Suppose bridges $b$ and $b'$ have exposure class spaces
\[
\mathcal{E}_b = \{\varepsilon_1, \dots, \varepsilon_i, \cdots, \varepsilon_{N_b}\}, 
\quad 
\mathcal{E}_{b'} = \{\varepsilon_1, \dots, \varepsilon_j, \cdots, \varepsilon_{N_{b'}}\},
\]
respectively. The corresponding joint exposure class space is defined as
$
\boldsymbol{\mathcal{E}} = \mathcal{E}_b \times \mathcal{E}_{b'},
$
where each joint class is given by
$
\boldsymbol{\varepsilon}_{ij} := (\varepsilon_i, \varepsilon_j).
$

Under realization $r$ ($r = 1, \dots, R$), let $\varepsilon_b^{(r)}$ and $\varepsilon_{b'}^{(r)}$ denote the sampled classes for bridges $b$ and $b'$, respectively, and define the joint observation:
$
\boldsymbol{\varepsilon}^{(r)} := \big(\varepsilon_b^{(r)}, \varepsilon_{b'}^{(r)}\big).$

The number of occurrences of joint class $(i,j)$ is estimated by
\[
N_{ij} = \sum_{r=1}^R \mathbf{1}\!\left(\boldsymbol{\varepsilon}^{(r)} = \boldsymbol{\varepsilon}_{ij}\right),
\]
and the corresponding empirical joint probability is given by
\[
\hat{\pi}(\varepsilon_i,\varepsilon_j) = \frac{N_{ij}}{R}.
\]

The mean of covariance $\mathbb{E}\left[\mathrm{Cov}(L_b, L_{b^\prime}|\mathcal{E})\right]$ is then estimated by enumerating over joint bridge class assignments for each bridge pair. Specifically, for a given bridge class pair $\boldsymbol{\varepsilon}_{ij}$, the conditional covariance $\mathrm{Cov}(L_b, L_{b^\prime}|\boldsymbol{\varepsilon_{ij}})$ is estimated, with its estimand denoted as $\sigma_{b,b^{\prime}|ij}$. Then, the mean of covariance is estimated by weighting these conditional covariances using the empirical joint class distribution $\hat{\pi}(\varepsilon_i,\varepsilon_j)$:

\begin{equation*}
    \hat{\mu}_{b|ij} = \frac{1}{N_{ij}}\sum_{r: \boldsymbol{\varepsilon}^{(r)} = \boldsymbol{\varepsilon}_{ij}} L_b^{(r)}
\end{equation*}

\begin{equation*}
    \hat{\mu}_{b^\prime|ij} = \frac{1}{N_{ij}}\sum_{r: \boldsymbol{\varepsilon}^{(r)} = \boldsymbol{\varepsilon}_{ij}} L_{b^\prime}^{(r)}
\end{equation*}

\begin{equation*}
    \hat{\sigma}_{b,b^\prime|ij}  = \frac{1}{N_{ij}-1}\sum_{r: \boldsymbol{\varepsilon}^{(r)} = \boldsymbol{\varepsilon}_{ij}} (L_b^{(r)}-\hat{\mu}_{b|ij})\cdot(L_{b^\prime}^{(r)}-\hat{\mu}_{{b^\prime}|ij})
\end{equation*}

\begin{equation*}
    \hat{\mu}[{\hat{\sigma}_{b,b^\prime|\mathcal{E}}}] = \sum_{i,j}\hat{\pi}(\varepsilon_i,\varepsilon_j) \cdot \hat{\sigma}_{b,b^\prime|ij} 
\end{equation*}

The exposure information variance  in \autoref{eq:total_var_regional} can be written as: 
\begin{equation*}
\begin{split}
    \mathrm{Var}^{\text{e}}(RL) &= \mathrm{Var} \left(\mathbb{E} \left[\sum_{b=1}^BL_b | \mathcal{E} \right] \right)  \\
    &= \mathrm{Var}\left(\sum_{b=1}^B \mathbb{E}[L_b|\mathcal{E}] \right)\\
    &= \sum_{b=1} ^ B\mathrm{Var}(\mathbb{E}[L_b|\mathcal{E}]) + 2\sum_{b<b^\prime}\mathrm{Cov}(\mathbb{E}[L_b|\mathcal{E}], \mathbb{E}[L_{b^\prime}|\mathcal{E}])
\end{split}
\end{equation*}
Similar to $\mathrm{Var}^{\text{base}}(RL)$, the first part of $\mathrm{Var}^{\text{e}}(RL)$ accounts for the sum of exposure information variance from individual bridges, and has been introduced in the main text. 
The second part is the covariance of inter-bridge conditional means. This captures the correlation between losses from bridge pairs. It is estimated by first computing class-specific conditional means for bridge $b$ and $b^\prime$ (i.e., ${\mu}_{b|i}$ and ${\mu}_{b^\prime|j}$) and then estimating the covariance of these conditional means.
\begin{equation*}
    \hat{\mu}_{b|i} = \frac{1}{N_i} \sum_{r: \varepsilon_b^{(r)} = \varepsilon_i} L_b^{(r)}
\end{equation*}

\begin{equation*}
    \hat{\mu}_{b^\prime|j} = \frac{1}{N_j} \sum_{r: \varepsilon_{b^\prime}^{(r)} = \varepsilon_j} L_{b^\prime}^{(r)}
\end{equation*}

\begin{equation*}
    \hat{\mu}_{b} = \sum_i \hat{\pi}(\varepsilon_i)\hat{\mu}_{b|i}
\end{equation*}

\begin{equation*}
    \hat{\mu}_{b^\prime} = \sum_j \hat{\pi}(\varepsilon_j)\hat{\mu}_{b^\prime|j}
\end{equation*}

\begin{equation*}
    \hat{\sigma}(\hat{\mu}_{b|i},\hat{\mu}_{b^\prime|j}) = \sum_{i,j} \hat{\pi}(\varepsilon_i,\varepsilon_j) \cdot (\hat{\mu}_{b|i}-\hat{\mu}_b) \cdot (\hat{\mu}_{b^\prime|j}-\hat{\mu}_{b^\prime})
\end{equation*}

\end{appendices}

\bibliographystyle{unsrt}  

\bibliography{sample}

\end{document}